\documentclass{article}
\usepackage{amsmath,amsfonts,amsthm}
\usepackage{natbib}
\usepackage{hyperref}
\usepackage{graphicx}
\usepackage{thmtools,mathtools,thm-restate,bbm}
\usepackage{subcaption}

\newtheorem{lemma}{Lemma}

\newcommand{\Prob}[1]{\ensuremath{\mathrm{Pr}\left( #1 \right)}}
\newcommand{\Expect}[1]{\ensuremath{\mathbb{E}\left[ #1 \right]}}
\newcommand{\Cov}[1]{\ensuremath{\mathrm{Cov}\left[ #1 \right]}}
\newcommand{\Var}[1]{\mathrm{Var}\left[ #1 \right]}
\newcommand{\indep}{\rotatebox{90}{\ensuremath{\models}}}

\newcommand{\distiid}{\overset{iid}{\sim}}
\newcommand{\ignore}[1]{}

\begin{document}
\title{Estimating Causal Peer Influence in Homophilous Social Networks by
Inferring Latent Locations}
\author{Edward McFowland {III} \thanks{Department of Information and Decision
Sciences, Carlson School of Management, University of Minnesota} \and Cosma
Rohilla Shalizi\thanks{Statistics Department, Carnegie Mellon University, and
the Santa Fe Institute}}

\date{}
\maketitle

\begin{abstract}
  Social influence cannot be identified from purely observational data on social
  networks, because such influence is generically confounded with latent
  homophily, i.e., with a node's network partners being informative about the
  node's attributes and therefore its behavior. If the network grows according
  to either a latent community (stochastic block) model, or a continuous latent
  space model, then latent homophilous attributes can be consistently estimated
  from the global pattern of social ties. We show that, for common versions of
  those two network models, these estimates are so informative that controlling
  for estimated attributes allows for asymptotically unbiased and consistent
  estimation of social-influence effects in linear models. In particular, the
  bias shrinks at a rate which directly reflects how much information the
  network provides about the latent attributes. These are the first results on
  the consistent non-experimental estimation of social-influence effects in the
  presence of latent homophily, and we discuss the prospects for generalizing
  them.
\end{abstract}


\section{Introduction: Separating Homophily from Social Influence}
It is an ancient observation that people are influenced by others (nearby) in
their social network---that is, the behavior of one node in a social network
adapts or responds to that of neighboring nodes. Such social influence is not
just a curiosity, but of deep theoretical and empirical importance across the
social sciences. It is also of great importance to various kinds of social
engineering, e.g., marketing (especially, but not only, ``viral'' marketing),
public health (over-coming ``peer pressure'' to engage in risky behaviors, or
using it to spread healthy ones), education (``peer effects'' on learning),
politics (``peer effects'' on voting), etc. Conversely, it is an equally ancient
observation that people are not randomly assigned their social-network
neighbors. Rather, they {\em select} them, and tend to select as neighbors those
who are already similar to themselves. (This is not necessarily because they
{\em prefer} those who are similar; all more-desirable potential partners might
have already been claimed or otherwise excluded
\citep{JLMartin-on-social-structures}.)  This {\bf homophily} means that network
neighbors are informative about latent qualities a node possesses, providing an
alternative route by which a node's behavior can be predicted from their
neighbors. Efforts to separate homophily from influence have a long history in
studies of networks \citep{Leenders-structure-and-influence}. Motivated by the
controversy over \citet{Christakis-Fowler-spread-of-obesity},
\citet{Homophily-contagion-confounded} showed that unless {\em all} of the nodal
attributes which are relevant to both social-tie formation {\em and} the
behavior of interest are observed, then social-influence effects are generally
unidentified. The essence of this result is that a social network is a machine
for {\em creating} selection bias\footnote{A turn of phrase gratefully borrowed
from Ben Hansen.}.

\citet[\S 4.3]{Homophily-contagion-confounded} did hint at a possible approach
for identification of social influence, even in an homophilous network. When a
network forms by homophily, a node is likely to be similar to its neighbors.
Following this logic, these neighbors are likely to be similar to {\em their}
neighbors and therefore the original node. In the simplest situations, where
there are only a limited number of node types, this means that a homophilous
network should tend to exhibit clusters with a high within-cluster tie density
and a low density of ties across clusters. Breaking the network into such
clusters might, then, provide an observable proxy for the latent homophilous
attributes. The same idea would work, {\em mutatis mutandis}, when those
attributes are continuous. \citet{Homophily-contagion-confounded} therefore
conjectured that, under certain assumptions on the network-growth process (which
they did not specify), unconfounded causal inferences could be obtained by
controlling for {\em estimated} locations in a latent space. Subsequently,
\citet{Davin-Gupta-Piskorski-separating-homophily} and \citet{Worrall-homophily}
showed that, in limited simulations, such controls can indeed reduce the bias in
estimates of social influence, at least when the network grows according to
certain, particularly well-behaved, models.

In this paper, we complement these simulation studies by establishing sufficient
conditions under which controlling for estimated latent locations leads to {\em
asymptotically} unbiased and consistent estimates of social-influence effects.
Additionally, we show that for a particular class of network models, the
remaining finite-sample bias shrinks exponentially in the size of the network,
while this bias shrinks polynomially for a more general class of network models.
To the best of our knowledge, our results provide the first {\em theoretical}
guarantees of consistent estimation of social-influence effects from
non-experimental data, in the face of latent homophily. Additionally, we provide
our own simulations to support and explore our theoretical results.

Section \ref{sec:setting} lays out the basics of our setting, starting with
assumptions about the processes of network formation and social influence (and
the links between them), and rehearsing relevant results from the prior
literature on latent community models (\S \ref{sec:communities}) and continuous
latent space models (\S \ref{sec:continuous-space}). Section
\ref{sec:community-control} presents our main results about the asymptotic
estimation of social influence in the presence of latent homophily (proofs are
deferred to \S \ref{sec:proofs}). Section \ref{sec:simulations} provides a set
of simulations that confirm our theoretical results and explore settings that
diverge from ours. Section \ref{sec:discussion} discusses the strengths and
limits of our results in the context of the related literature.

\section{Setting and Assumptions}
\label{sec:setting}
\begin{figure}
  \includegraphics[width=\textwidth]{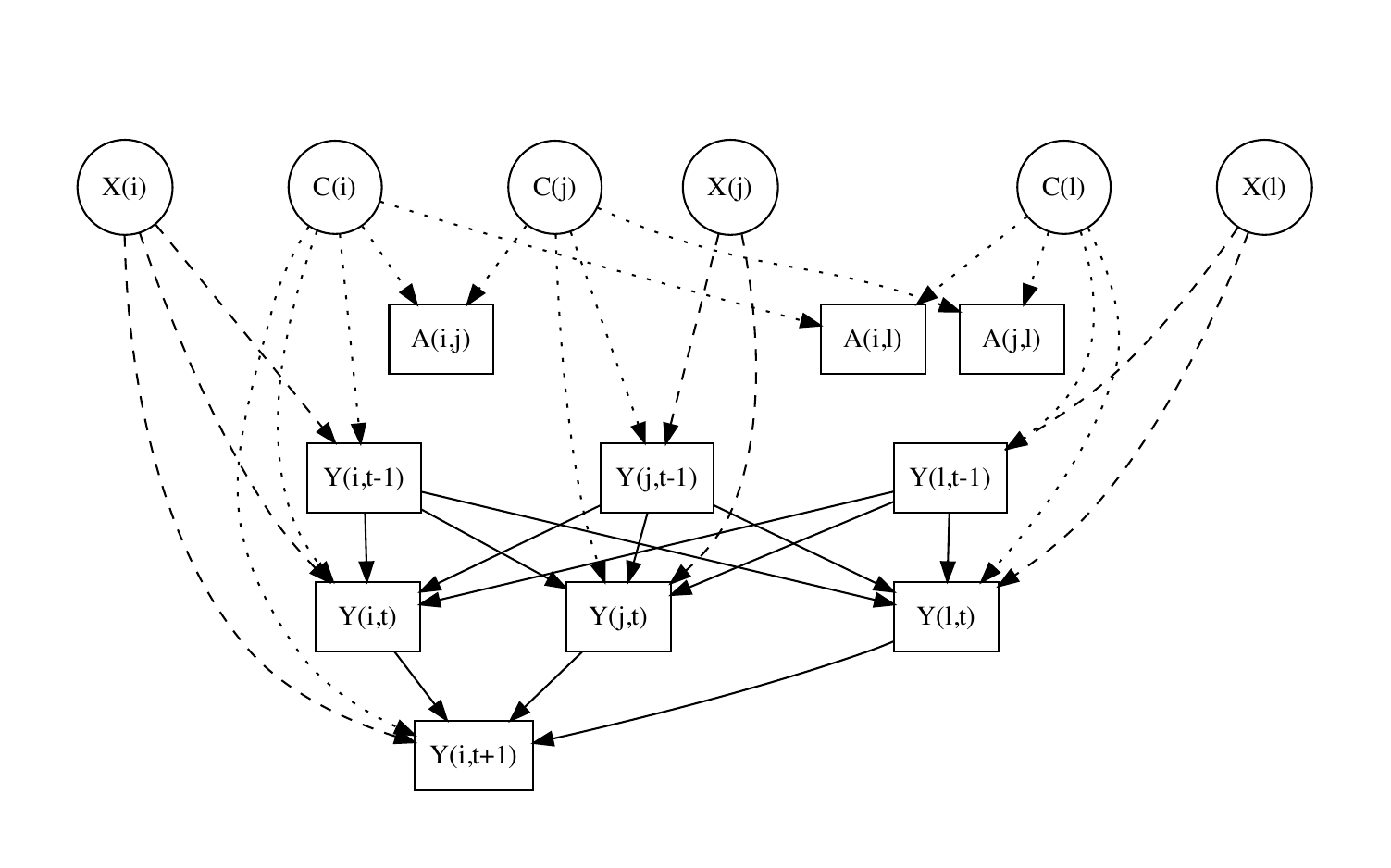}
  \caption{The graphical causal model for our setting. Boxes indicate
  observables, and circles latent variables; solid lines indicate causal
  relations between observables (either autoregressive or peer-influence), while
  dotted lines indicate the influence of latent homophilous variables, and
  dashed lines indicate the influence of other covariates. For simplicity, we
  omit $Y(j, t+1)$ and $Y(l, t+1)$, as well as their associated arrows.}
  \label{fig:graphical-model}
\end{figure}

The graphical causal model\footnote{We do not mean to take sides in the dispute
between the partisans of graphical causal models and those of the
potential-outcomes formalism. The expressive power of the latter is strictly
weaker than that of suitably-augmented graphical models
\citep{Richardson-Robins-SWIGs}, but we could write everything here in terms of
potential outcomes, albeit at some cost in space and notation.} capturing social
influence in our setting is shown in Figure \ref{fig:graphical-model}. More
specifically, we are interested in the patterns of a certain behavior or outcome
over time, across a social network of $n$ nodes. The behavior of node $i \in
\{1, \ldots, n\}$ at time $t \in \{1, \ldots, T\}$ is observed and represented
by random variable $Y_{i,t} \in \mathbb{R}$, for some given time-horizon $T$.
Social network ties (or links) are also observed and represented through an $n
\times n$ adjacency matrix $A$, with $A_{ij} = 1$ if $i$ receives a tie from
$j$, and $A_{ij} = 0$ otherwise. In many contexts these ties are undirected, so
$A_{ij} = A_{ji}$, but generally our results do not require this. (In the latent
community setting [\S \ref{sec:communities}], the procedure considered by
\citet{Gao-sbm-minimax_algo} assumes an undirected network, and therefore the
results of ours which rely on that procedure also make this assumption.) As this
notation suggests, we assume that the network of social ties does not change, at
least over the time-scale of the observations\footnote{Latent space modeling of
dynamic networks is still in its infancy. For some preliminary efforts, see,
e.g., \citet{DuBois-Butts-Smyth-blockmodeling-of-dynamics,
Ghasemian-et-al-community-in-dynamic-networks} for block models, and
\citet{Sarkar-Moore-dynamical-social-network} for continuous-space models.}.

In addition to the observed behaviors and ties of node $i$, we assume there
exist a $d$-dimensional latent vector $C_i$ which controls its location in the
network; we define $C$ as the array $[C_1, C_2, \ldots, C_n]$. Furthermore, we
assume that $\Prob{A_{ij}=1|C} = w(C_i, C_j)$ for some measurable function $w$,
and that the random variables $A_{ij}$ and $A_{lm}$ are conditionally
independent given $C$, $\forall~i,j \ne l,m$. The time-invariant vector $X_i$
represents the set of all other (i.e., network irrelevant) attributes for node
$i$, which effect $Y_{i,t}$ but not $A_{ij}$.

The linear structural-equation model that explains the behavior of node $i$ at
time $t$ is thus
\begin{equation}
	Y_{i,t+1} = \alpha_0 + \alpha_1 Y_{i,t} +
	\beta\frac{\sum_{j}{\left(Y_{j,t}A_{ij}\right)}}{\sum_{j}{A_{ij}}} +
	\gamma_1^T C_{i} + \gamma_2^T X_{i} + \epsilon_{i,t+1}
	\label{eqn:structural-equation} ~,
\end{equation}
where $\gamma_1$ and $\gamma_2$ serve as appropriately-sized
vectors of coefficients. Under the assumptions of linearly-independent
regressors and strict exogeneity---i.e., $\Expect{\epsilon_{i,t+1} | Y_{i,t},
A_{ij}, C_{i}, X_{i}} = 0 ~ \forall i,j,t$---our goal is to identify, and
estimate, $\beta$, the coefficient for social influence. Given that we neither
observe $X_i$ nor $C_i$, we cannot estimate the regression coefficients in a
model of the form presented in \eqref{eqn:structural-equation}. However, we {\em
can} estimate the coefficients of the following model:
\begin{equation}
  Y_{i,t+1} = \alpha_0 + \alpha_1 Y_{i,t} +
  \beta\frac{\sum_{j}{\left(Y_{j,t}A_{ij}\right)}}{\sum_{j}{A_{ij}}} +
  \gamma_0^T \hat{C}_i + \eta_{i,t+1} \label{eqn:effective-model} ~,
\end{equation}
where $\hat{C}_i$ is an estimated or discovered location for node $i$ and the
noise term $\eta_{i,t+1}$ can be defined as
\begin{equation*}
  \eta_{i,t+1} = \epsilon_{i,t+1} + \gamma_2^T X_{i} + \left(\gamma_1^T C_{i} -
  \gamma_0^T \hat{C}_i\right) ~.
\end{equation*}
Our general setting is therefore defined by an additional assumption:
\begin{equation}
  X_{i} \indep Y_{j, t} | \hat{C}_i
  \label{eqn:screening-egos-attributes-from-alters-behavior-2}.
\end{equation}
\ignore{Note that implicitly $X_{i} \indep Y_{j, t}$ is assumed by our causal
graph model (Figure \ref{fig:graphical-model}) and structural-equation model
\eqref{eqn:structural-equation}. }The assumptions of independence must be
justified on substantive grounds, in the specific context of the study where
social influence is being estimated.

\subsection{Discussion on the General Setting}
\label{sec:setting-discussion}
We find it beneficial to provide intuition on how various facets of our setting
enable the identification of social-influence effects. We begin by recognizing
that the relatively permanent attributes of node $i$ can be divided in two
cross-cutting ways. On the one hand, some attributes are (in a given study)
observable or manifest, and others are latent. On the other hand, a given
attribute could be a cause of the behavior of interest $Y_{i,t}$, or a cause of
network ties ($A_{ij}$), or of both. (Attributes which are irrelevant to both
behavior and network ties are ignored here as they have no bearing on our
ultimate goal).

One of the key assumptions embedded in our tie formation process (i.e.,
$\Prob{A_{ij}=1|C} = w(C_i, C_j)$) is that {\em all} of the network-relevant
attributes of node $i$ can be represented by a single vector-valued latent
variable $C_i$, whether or not they are also relevant to the behavior of
interest. There may be attributes that are incorporated into $C_i$ which are
relevant {\em only} to network ties, not behavior, and independent of the other
attributes; these are of no concern to us, and can be regarded as part of the
noise in the tie-formation process. Network models that satisfy this
assumption---i.e., that all ties are conditionally independent of each other
given the latent variables for each node---are sometimes called ``graphons'' or
``$w$-random graphs'' and are clearly exchangeable (permutation-invariant) over
nodes. Conversely, the Aldous-Hoover theorem \citep[ch.
7]{Kallenberg-symmetries} shows that this condition is, in fact, the generic
form of exchangeable random networks. Our subsequent assumption, speaking
roughly, is that by observing the whole network $A_{ij}$ (which inherently
includes the information it contains with respect to the latent array $C$),
$Y_{i,t}$ provides no additional information (in the limit) for node $i$'s
latent location $C_i$.

We also recognize that as a result of the assumptions of linear-dependence and
strict exogeneity in \eqref{eqn:structural-equation}, if all the variables
relevant to tie-formation and node behavior are observed, the ordinary least
squares (OLS) estimator provides an unbiased estimate ($\hat{\beta}_{OLS}$) of
$\beta$. However, since $C_i$ and $X_i$ are both unobserved, and therefore their
effects are contained in the $\eta_{i,t+1}$ of \eqref{eqn:effective-model},
$\hat{\beta}_{OLS}$ will generally contain omitted variable bias if either of
these latent variables are correlated with $Y_{j,t}$, conditional on the
observed regressors. Intuitively, the latent nature of $X_i$ will not produce
bias because \eqref{eqn:screening-egos-attributes-from-alters-behavior-2}
implies that given estimated locations, nothing can be learned about a node's
unobserved, network-irrelevant attributes by observing a neighbor's behavior (or
vice-versa). Mathematically, this means that the contribution of $\gamma_2^T
X_{i}$ to $\eta_{i,t+1}$ is uncorrelated with $Y_{j, t}$, given the estimated
locations, and therefore this term does not bias the estimates of $\beta$;
instead it just increases the variance of the noise term. It is also not
necessary that $E[\eta_{i,t+1}] = 0$; if it has a non-zero value, it would then
be incorporated into the estimate of the intercept ($\alpha_0$), and therefore
not induce bias in $\beta$. We have therefore to only consider the other
contribution to $\eta_{i,t+1}$, $(\gamma_1 C_i - \gamma_0 \hat{C}_i)$, and
whether it is correlated with $Y_{j, t}$ given $\hat{C}_i$ and $Y_{i, t}$.

$\lVert C_i - \hat{C}_i\rVert$ is the error in estimating the true location,
which manifests as measurement error in OLS estimation of $\beta$ in
\eqref{eqn:effective-model}. Given that $Y_{j, t}$ is a causal descendant of
$C_j$, and $C_j$ is positively correlated with $C_i$ if $A_{ij}=1$ (from
homophily), this measurement error induces bias in the estimate of $\beta$.
Intuitively (and formally shown in Lemma \ref{lemma:no-error-no-problem} below)
if $\hat{C} = C$ (i.e., there is no measurement error) the OLS estimate of
$\beta$ in \eqref{eqn:effective-model} will be unbiased and consistent;
although, this estimate will likely have a larger variance than would the OLS
estimate of $\beta$ from \eqref{eqn:structural-equation}, given that the former
estimate does not control for $X_{i}$. It should further be plausible (and is
formally shown in \S \ref{sec:community-control} below) that if $\hat{C}$ is a
``good enough'' estimate of $C$---i.e., one which is consistent and converges
sufficiently rapidly---the covariance between $\eta_{i,t+1}$ and $Y_{j, t}$
shrinks fast enough that the OLS estimate from \eqref{eqn:effective-model} will
still yield asymptotically unbiased and consistent estimates of $\beta$.
Essentially, the OLS estimator for $\beta$ in \eqref{eqn:effective-model}
trades-off the bias (experienced by the OLS estimator for $\beta$ in
\eqref{eqn:structural-equation}) from omitting the latent location variable
$C_i$, with the bias from measuring (estimating) the location imprecisely with
$\hat{C}_i$. However, the ability to obtain a ``good enough'' estimate of $C$
will make this trade-off worthwhile; if the measurement error converges to zero,
then the bias it induces should also converge to zero, while the omitted
variable bias persists.

There does not (yet) exist results providing such ``good enough'' estimates of
latent node locations $\hat{C}$ for arbitrary graphons. For this reason, our
results specialize to two settings, where the latent node locations $C$ and the
link-probability function $w$ take particularly tractable forms: latent
community (stochastic block) models and the more general (continuous) latent
space models. Both model types have been extensively explored in the literature.
It is by building on results for these models that we can find regimes where the
social-influence coefficients can be estimated consistently. It is, however,
worth noting that for any graphon model where ``good enough'' estimates of
latent node locations $\hat{C}$ exist, an analog to our results for latent space
models (Theorem \ref{asymptotic-consistent-for-latent-space}) can be built.

\subsection{The Latent Communities Setting}
\label{sec:communities}
In our first setting, we presume that nodes split into a finite number of
discrete types or classes ($k$), which in this context are called {\bf blocks},
{\bf modules} or {\bf communities}. More precisely, there exists a function
$\sigma \colon \{1,\dots, n\} \mapsto \{1,\dots, k\}$ assigning nodes to
communities. We specifically assume that the network is generated by a {\bf
stochastic block model}, which is to say that there are $k$
communities\footnote{Some of the theory we rely on below allows the number of
communities to grow with the size of the network, though with at a rate posited
to be known {\em a priori}, and not too fast. We leave dealing with this
complication to future work.}, that $\sigma(i) \distiid \rho$, for some fixed
(but unknown) multinomial distribution $\rho$, and that $w$ is given by a
$k\times k$ {\bf affinity matrix}, so that
\[
\Prob{A_{ij}=1|\sigma(i)
  = a, \sigma(j) = b} = w_{ab} ~.
\]
We may translate between $\sigma$ (a sequence of categorical variables) and our
earlier $C$ (an $n \times d$ matrix of node locations) by the usual device of
introducing indicator or ``dummy'' variables for $k-1$ of the communities, so
that $C_i$ is a $k-1$ binary vector (i.e., $d = k-1$) which is a function of
$\sigma(i)$ and vice versa. Each possible value of $C_i$ is either the origin,
or a corner of the simplex; this basic observation will be important below.

The objective of community detection or community discovery is to provide an
accurate estimate $\hat{\sigma}$ or $\hat{C}$ from the observed adjacency matrix
$A$, i.e., to say which community each node comes from, subject to a permutation
of the label set. (``Accuracy'' here is often measured as the proportion of
mis-classification.)  Since the problem was posed by
\citet{Girvan-MEJN-community-structure} a vast literature has emerged on the
topic, spanning many fields, including physics, computer science, and
statistics; see \citet{Fortunato-community-detection} for a review. However, we
may summarize the most relevant findings as follows.

\begin{enumerate}
\item For networks which are generated from latent community models, under very
  mild regularity conditions, it is possible to recover the communities
  consistently, i.e., as $n \rightarrow \infty$, $\Prob{\hat{C} \neq C}
  \rightarrow 0$ \citep{Bickel-Chen-on-modularity,
    Zhao-Levina-Zhu-consistency-of-community-detection}. That is, with
  probability tending to one, {\em all} of the community assignments are
  correct, up to a global permutation of the labels between $C$ and $\hat{C}$.
\item Such consistent community discovery can be achieved by algorithms whose
  running time is polynomial in $n$.
\item The minimax rate of convergence is in fact exponential in $n$ (and can be
  achieved by the algorithms mentioned below).
\end{enumerate}
These points, particularly the last, will be important in our argument below,
and so we now elaborate on them.

Recently, \citet{Zhang-sbm_minimax} proved that under very mild regularity
conditions the minimax rate of convergence for undirected networks generated
from latent community models is in fact exponential in $n$. Furthermore,
\citet{Gao-sbm-minimax_algo} exploits techniques provided by
\citet{Zhang-sbm_minimax} to propose an algorithm polynomial in $n$ that
achieves this minimax rate, under slightly modified but equally mild regularity
conditions. More precisely, \citet{Gao-sbm-minimax_algo} considers a general
undirected stochastic block model, parametrized by $n$, the number of nodes;
$k$, the number of communities; $a$ and $\alpha \ge 1$, where $\frac{a}{n} =
\min_{i}{w(i,i)} \le \max_{i}{w(i,i)} \le \frac{\alpha a}{n}$, ensuring that
within-community edges are ``sufficiently'' dense; $b$, where $\frac{b
\alpha}{n} \le \frac{1}{k(k-1)} \sum_{i \ne j}{w(i,j)} \le \max_{i \ne
j}{w(i,j)} \frac{b}{n}$, with $0 < \frac{b}{n} < \frac{a}{n} < 1$, ensuring that
between-community edges are ``sufficiently'' sparse; and $\beta \ge 1$, where
the number of nodes in community $k$, $n_k \in \left[\frac{n}{\beta k},
\frac{\beta n}{k} \right]$, ensuring that community sizes are ``sufficiently''
comparable. \citet{Zhang-sbm_minimax} and \citet{Gao-sbm-minimax_algo} diverge
slightly as the former only requires $\max_{i \ne j}{w(i,j)} \le \frac{b}{n}$
and $\frac{a}{n} \le \min_{i}{w(i,i)}$. Additionally, the latter slightly
restricts the parameter space by requiring the $k^{th}$ singular value of the
affinity matrix $w$ to be greater than some parameter $\lambda$. The general
context of the theory described in \citet{Zhang-sbm_minimax,
Gao-sbm-minimax_algo} is defined for absolute constant $\beta \ge 1$ and also in
\citet{Gao-sbm-minimax_algo} for absolute constant $\alpha \ge 1$, while $k$,
$a$, $b$, and $\lambda$ are functions of $n$ and therefore vary as $n$ grows.
However, in our context, the network does not change (over the time-scale of
interest); therefore, we only consider latent communities where $k$,
$\frac{a}{n}$, $\frac{b}{n}$, and $\lambda$ are also absolute constants. We
shall refer to this whole set of restrictions on the latent community model as
``the GMZZ conditions''.

For a latent community model satisfying the GMZZ conditions, the minimax rate of
convergence for the expected {\em proportion} of errors is
\begin{align}
    \label{eq:converg-rate-2}
    \exp{\left(-(1+o(1))\frac{nI}{2}\right)}, & \quad k=2 \\
    \label{eq:converg-rate-3}
    \exp{\left(-(1+o(1))\frac{nI}{\beta k}\right)}, & \quad k \ge 3,
\end{align}
where $I$ is the \citet{Renyi-introduces-Renyi-entropy} divergence of order
$\frac{1}{2}$  between two Bernoulli distributions with success probabilities
$\left(\frac{a}{n}\right)$ and $\left(\frac{b}{n}\right)$:
$D_{\frac{1}{2}}\left(\text{Ber}\left(\frac{a}{n}\right) \|
\text{Ber}\left(\frac{b}{n}\right) \right)$. Recall that $\beta$ in addition to
$k$, $\frac{a}{n}$, $\frac{b}{n}$ are, in our context, constant in $n$;
therefore, \eqref{eq:converg-rate-2} and \eqref{eq:converg-rate-3} both reduce
to $\exp{\left(-O(n)\right)}$. The algorithm of \citet{Gao-sbm-minimax_algo}
achieves this rate at a computational cost polynomial in $n$. More specifically,
the time complexity of the algorithm is (by our calculations) at most $O(n^3)$,
but we do not know whether this is tight. It would be valuable (but beyond the
scope of this work) to know whether this rate is also a lower bound on the
computational cost of obtaining minimax error rates, and if the complexity could
be reduced in practice for very large graphs via parallelization.

We close this section by introducing a bit of notation (which will simplify some
later statements) and making a claim (which will be supported later). We will
write $\delta(n)$ for the error probability, i.e., the probability that
$\hat{C}_i \neq C_i$ for at least one $i \in 1:n$.\ignore{In contrast, $G_n$
will be the indicator variable for the complementary event, $G_n=1$ if
$\hat{C}_i=C_i$ for all $i\in 1:n$ and $G_n=0$ otherwise. (These choices will
simplify some later statements.)} The claim is that even though the results of
\citet{Zhang-sbm_minimax} and \citet{Gao-sbm-minimax_algo} concern the
proportion of mis-classified nodes, they actually constrain the probability of
making any mis-classifications at all, and imply $\delta(n) = e^{-O(n)}$ (Lemma
\ref{lemma:asymptotic-covariance-for-block-models}).

\subsection{The Continuous Latent Space Setting}
\label{sec:continuous-space}
The second setting we consider is that of continuous latent space models. In
this setting, the latent variable on each node, $C_i$, is a point in a
continuous metric space (often but not always $\mathbb{R}^d$ with the Euclidean
metric), and $w(C_i, C_j)$ is a decreasing function of the distance between
$C_i$ and $C_j$, e.g., a logistic function of the distance. This
link-probability function is often taken to be known {\em a priori}. The latent
locations $C_i \distiid F$, where $F$ is a fixed but unknown distribution, or,
more rarely, a point process. Different distributions over networks thus
correspond to different distributions over the continuous latent space, and vice
versa.

Parametric versions of this model have been extensively developed since
\citet{Hoff-Raftery-Handcock}, especially in Bayesian contexts. Less attention
has been paid to the consistent estimation of the latent locations in such
models, than to the estimation of community assignments in latent community
models. Recent results by \citet[ch. 3]{Asta-thesis}, however, show that when
$w$ is a smooth function of the metric whose logit transformation is bounded,
the maximum likelihood estimate $\hat{C}$ converges to $C$. Moreover, the
probability of an error of size $\epsilon$ or larger is $O(\exp{\left(-\kappa
\epsilon n^2\right)})$, where the constant $\kappa$ depends on the purely
geometric properties of the space (see \S
\ref{sec:continuous-latent-space-setting} below). This result holds across
distributions of the $C_i$, but may not be the best possible rate.

\section{Control of Confounding}
\label{sec:community-control}
Given the (assumed) true structural equation in \eqref{eqn:structural-equation},
our ultimate goal is to provide both an estimator of $\beta$, and the
corresponding sufficient conditions under which that estimator will have
desirable statistical properties. Recall that these properties of the estimator
are evaluated in the presence of estimated or discovered node locations
$\hat{C}$, rather than the true locations $C$. Going forward, therefore, unless
otherwise noted, our estimator of interest is OLS for $\beta$ in
\eqref{eqn:effective-model}. Finally, all proofs of the results stated below are
provided in \S\ref{sec:proofs}.

We begin by establishing this estimator's properties in a baseline case: when
the estimates of node locations are perfect, $\Prob{C \neq \hat{C}} = 0$.
\begin{restatable}{lemma}{noerror}
  \label{lemma:no-error-no-problem}
  Under the assumptions from Section \ref{sec:setting}, if $\Prob{C \neq
  \hat{C}} = 0$, then the ordinary least squares estimate of $\beta$ in
  \eqref{eqn:effective-model} is unbiased and consistent. \ignore{an unbiased
  and consistent estimate of the social-influence coefficient ($\beta$) from
  \eqref{eqn:structural-equation}.}
\end{restatable}

Given that we establish that OLS estimator will exhibit unbiasedness and
consistency, when node locations can be perfectly inferred, let us now consider
its properties when the node location are inferred with error. The covariance of
interest is that between $\frac{\sum_{j}{\left(Y_{j,t}A_{ij}\right)}}{\sum_{j}{
A_{ij}}}$ and the contribution to the error---i.e., $\eta_{i,t+1}$ in
\eqref{eqn:effective-model}---arising from using the estimated rather than the
real communities. We have seen (\S\ref{sec:setting-discussion}\ignore{page\
\pageref{error-term-from-c-vs-hat-c}}), that in our setting, under assumption
\eqref{eqn:screening-egos-attributes-from-alters-behavior-2}, this term is just
$\gamma_1^T C_i - \gamma_0^T \hat{C}_i$. Moreover, we will only need to consider
that covariance conditional on $\hat{C}_i$ and $\hat{C}_j$, and the other
regressor in  \eqref{eqn:effective-model}, i.e., $Y_{i, t}$.
\begin{restatable}{lemma}{altrsanderrors}
  \label{lemma:covariance-between-altrs-behavior-and-estimation-errors}
  Suppose that the assumptions from Section \ref{sec:setting} hold. Then
 \begin{align}
     &\Cov{\frac{\sum_{j}{\left(Y_{j,t}A_{ij}\right)}}{\sum_{j}{A_{ij}}} ,
     (\gamma_1^T C_i - \gamma_0^T \hat{C}_i) \middle| A, Y_{i,t}}
     \label{eqn:covariance-between-altrs-behavior-and-estimation-errors}\\
     &~=  \frac{\sum_{j}{A_{ij}\gamma_1^T \left( \Cov{C_i, C_j
     \middle| A} + \xi_{ij} \Var{C_i|A} + \sum_{l \neq i, j}{\zeta_{ijl}
     \Cov{C_i, C_l|A}}\right) \gamma_1}}{\sum_{j}{A_{ij}} \nonumber
    } ,
 \end{align}
   where by $\Cov{C_i, C_j}$ we mean the $d \times d$ matrix of coordinate-wise
   covariances,  and similarly for $\Var{C_i}$, and the $\xi$s and $\zeta$s are
   constants calculable in terms of the model coefficients and the adjacency
   matrix (and are made explicit in the proof of the lemma).
\end{restatable}

Lemma \eqref{lemma:covariance-between-altrs-behavior-and-estimation-errors}
establishes an important relationship between the bias experienced by the OLS
estimator and the degree of homophily in the network. Recall that we observe a
network (represented by $A$) whose ties are formed under homophily, based on
(unobserved) node locations. Furthermore, it is precisely the fact that this
network is observed (and conditioned on) that opens the confounding backdoor
pathway in the causal graph (Figure \ref{fig:graphical-model}). For clarity, the
network is conditioned on because it enables the true structural equation
\eqref{eqn:structural-equation} to select if the behavior of node $i$ is
regressed on the behaviors of nodes $j$. Moreover, when homophily has a large
impact on tie formation, the value $\Cov{C_i, C_j}$ will be large, as node $i$
will have more connections from closer nodes $j$ (i.e., roughly, $C_i \approx
C_j$). We then recognize that although observing the network $A$ (and failing to
observe the latent locations $C$) opens the homophilous confounding pathway, the
network also manifests this homophily in the ties that are formed, which can be
used to form estimates of the latent locations $\hat{C}$. Further, conditioning
on $A$ implies conditioning on $\hat{C}_i$ and $\hat{C}_j$, as they are
deterministic (albeit complicated) functions of $A$. From Lemma
\eqref{lemma:covariance-between-altrs-behavior-and-estimation-errors} we then
see that the bias in our estimate $\hat{\beta}_{OLS}$ is proportional to the
amount of covariance between the true latent locations of nodes that share a
tie, beyond that which is accounted for by their location estimates. Therefore,
when this conditional covariance is zero, $\hat{\beta}_{OLS}$ is unbiased and
consistent.

There are two individually sufficient (but not necessary) conditions for
\eqref{eqn:covariance-between-altrs-behavior-and-estimation-errors} to be zero:
\begin{enumerate}
  \item $C_i \indep C_j | \hat{C_i},\hat{C_j}$, i.e., $C_i$ and $C_j$ are
    independent given their estimates,
  \item $C_i = \hat{C_i}$ and $C_j = \hat{C_j}$, i.e., $C_i$ and $C_j$ are equal
    to their estimates.
\end{enumerate}
The second condition will generally not be true at any finite $n$. The first
condition is also very strong; it implies that $\hat{C}$ is (roughly speaking) a
sufficient statistic for $C$. This sufficiency property implies that even in
learning $C_i$ (the true location of node $i$) we obtain no additional
information about $C_j$ (the location of any other node $j$) not already
captured in $\hat{C}$. We are not aware of any estimates of latent node
locations in network models which have such a sufficiency property, and we
strongly suspect this is because they generally are {\em not} sufficient. (To
get a sense of what would be entailed, suppose that $A_{ij}=1$, and we knew we
were dealing with a homophilous latent community model. Then $\hat{C}$ would
have to be so informative that even if an Oracle told us $C_i$, our posterior
distribution over $C_j$ would be unchanged.)  We may, however, make further
progress in the two specific settings of latent communities and of continuous
latent spaces.

\subsection{Control of Confounding with Latent Communities}
Let us first consider the setting where the network formation is that of a
homophilous latent community process, which follows the conditions laid out in
\S \ref{sec:communities}. In such a setting, we can make additional statements
with respect to the $\Cov{C_i, C_j \middle| \hat{C}_i, \hat{C}_j}$, and
subsequently the bias experienced by the OLS estimator. More specifically, these
statements are made assuming a deterministic and minimax algorithm---one that
achieves the minimax rate of convergence for the expected proportion of node
location errors ---is utilized to estimate $\hat{C}$ as in
\citet{Gao-sbm-minimax_algo}.
\begin{restatable}{lemma}{asycovblocks}
  \label{lemma:asymptotic-covariance-for-block-models}
  Suppose that the assumptions from Section \ref{sec:setting} hold, the network
  forms according to a latent community model, satisfying the GMZZ conditions,
  and $\hat{C}$ is estimated using a minimax algorithm. Then
  \begin{equation*}
    \Prob{\sum_i^n{\mathbbm{1}\{\hat{C}_i \ne C_i\}}  \geq 1} \le e^{-O(n)}.
  \end{equation*}
\end{restatable}

We therefore have that the probability of making any error in the estimation of
latent node locations converges (exponentially) to zero in $n$. This result from
Lemma \ref{lemma:asymptotic-covariance-for-block-models} will play a critical
role in proving the next result:
\begin{restatable}{lemma}{fincovblocks}
  \label{lemma:covariance-for-block-models}
  Suppose that the assumptions from Section \ref{sec:setting} hold, the network
  forms according to a latent community model, and $\hat{C}$ is estimated by a
  deterministic algorithm with error rate $\delta(n)$. Then
\begin{equation*}
  \Cov{C_i, C_j \middle| A} = O(\delta(n)).
\end{equation*}

If, in addition, the latent community model satisfies the GMZZ equations, and
$\hat{C}$ is estimated using a minimax algorithm, then
\begin{equation*}
  \Cov{C_i, C_j \middle| A} = O\left(e^{-O(n)}\right).
\end{equation*}
\end{restatable}

The ability to not only show the convergence of $\Cov{C_i, C_j \middle| A}$, but
also its rate of decay for finite-$n$ leads to a number of important
conclusions.
\begin{restatable}{theorem}{thmblocks}
  \label{theorem:asymptotic-consistent-for-block-models}
  Suppose that the assumptions from Section \ref{sec:setting} hold, the network
  forms according to a latent community model, and $\hat{C}$ is estimated with
  error rate $\delta(n)$. Then the ordinary least squares estimate for $\beta$
  in \eqref{eqn:effective-model} is asymptotically unbiased and consistent, and
  the pre-asymptotic bias is $O(\delta(n))$. If, in addition, the latent
  community model satisfies the GMZZ conditions and $\hat{C}$ is estimated using
  a deterministic and minimax algorithm, then the pre-asymptotic bias is
  exponentially small in $n$.
\end{restatable}
We suspect that it is also possible to provide a precise expression of a
deterministic finite-$n$ bound on the bias---likely as the solution to an
optimization problem involving (unknown) parameters of the structural
equation~\eqref{eqn:structural-equation}---but leave this as a useful topic for
future investigation.

{\em Note:} We have stated Lemma \ref{lemma:covariance-for-block-models} and
Theorem \ref{theorem:asymptotic-consistent-for-block-models} (and the subsidiary
Lemma \ref{lemma:label-variance-for-block-models}) in two parts to clarify that
most of their logic will apply whenever {\em some} deterministic algorithm is
capable of community discovery with a vanishing error rate $\delta(n)$. The GMZZ
conditions are invoked as regularity conditions under which $\delta(n)$ can be
made exponentially small at only a polynomial computational cost. If the GMZZ
conditions are implausible for a particular application, but some other
algorithm can, in that situation, deliver $\delta(n) \rightarrow 0$, then it can
be used instead within the scope of our analysis.

\subsection{Control of Confounding with Continuous Latent Space}
\label{sec:continuous-latent-space-setting}
We now turn our attention to setting where the network follows a homophilous
continuous latent space model. Recall that our treatment of the latent
\emph{community} setting relies on the fact that $\Prob{\hat{C} \neq C}
\rightarrow 0$, i.e., with probability tending to one the estimated communities
match the actual communities exactly. Importantly, this is not known to happen
for continuous latent space models, and seems very implausible for estimates of
continuous quantities, however we still can make progress.

As mentioned in \S \ref{sec:continuous-space}, \citet[ch.\ 3]{Asta-thesis} has
shown that if the link-probability function is known and has certain natural
regularity properties (detailed below), then the probability that the {\em sum}
of the distances between true locations and their maximum likelihood estimates
exceeds $\epsilon$ goes to zero exponentially in $\epsilon n^2$ (at least). More
specifically, the result requires the link-probability function to be smooth in
the underlying metric and bounded on the logit scale, and requires the latent
space's group of isometries\footnote{An isometry is a transformation of a metric
space which preserves distances between points. These transformations naturally
form groups, and the properties of these groups control, or encode, the geometry
of the metric space \citep{Brannan-Esplen-Gray}.} to have a bounded number of
connected components. (This is true for Euclidean spaces of any finite
dimension, where the number of connected components is always $2$.) If these
above conditions are met---which we shall refer to as ``the Asta
conditions''---then
\begin{equation*}
  \Prob{\sum_{i=1}^{n}{d(\hat{C}_i, C_i)} \geq \epsilon} \leq \mathcal{N}(n,
  \epsilon) e^{-\kappa n^2 \epsilon}
\end{equation*}
where the $\mathcal{N}$ is a known function, polynomial in $n$ and in
$1/\epsilon$, depending only on the isometry group of the metric, and $\kappa$
is a known constant, calculable from the isometry group and the bound on the
logit. Since the maximum of $n$ distances is at most the sum of those distances,
this further implies that
\begin{equation}
  \Prob{\max_{i\in 1:n}{d(\hat{C}_i, C_i)} \geq \epsilon)} \leq \mathcal{N}(n,
  \epsilon) e^{-\kappa n^2 \epsilon}.
  \label{eqn:asta-bound}
\end{equation}
With this, we can make the following asymptotic result.
\begin{restatable}{theorem}{thmspace}
  \label{asymptotic-consistent-for-latent-space}
  Suppose that the assumptions from Section \ref{sec:setting} hold, the network
  forms according to a continuous latent space model satisfying the Asta
  conditions, that the node-location distribution $F$ has compact support, and
  that $\hat{C}$ is estimated by maximum likelihood. Then the ordinary least
  squares estimate for $\beta$ in \eqref{eqn:effective-model} is asymptotically
  unbiased and consistent, and the pre-asymptotic bias is polynomially small in
  $n$.
\end{restatable}
The Asta conditions do not require $F$ to have compact support, but we use this
assumption for mathematical convenience in our derivation of the bound on the
bias. The assumption does, strictly speaking, rule out using a Gaussian
distribution for the latent locations. It is, however, compatible with using a
Gaussian that is truncated to 0 beyond some (large) distance from the origin. We
suspect the compact-support assumption can be weakened to merely assuming that
$F$ is tight, or that it has sufficiently light tails, but leave this to future
work. We suspect that it is also possible to provide a precise expression of a
deterministic finite-$n$ bound on the bias---though likely not the solution to
optimization problem, as we suspect for latent community models---and leave this
too as a useful topic for future investigation.

\section{Simulations}
\label{sec:simulations}
In observational studies over social networks, consistent estimation of the
social-influence parameter requires the ability to disentangle its effect from
that of homophily. Above, we gave conditions under which consistent (and {\em
asymptotically} unbiased) estimates of social influence is possible. The
simulations here aim to provide an empirical complement to these theoretical
results, verifying that our approach does in fact provide consistent estimates
of peer-influence, and achieves relatively small amounts of bias even at
manageable sample sizes. Additionally, we explore how estimates of the
peer-influence parameter behave as we (smoothly) depart from the conditions of
our theory, confirming that the results are robust to at least some violations
of the assumptions. Finally, the evaluation of our approach in these simulations
are done in the context of other estimation approaches, for proper comparisons.

\subsection{Simulation Setup}
Given that \citet{Davin-Gupta-Piskorski-separating-homophily} has already
conducted an empirical simulation study in the context of latent space models,
we will consider the latent community model setting to investigate our
theoretical results via simulation. We use the following \emph{R}~\citep{R}
packages to build our simulated network models: \emph{hergm}
\citep{Schweinberger-hergm}, \emph{mlergm} \citep{Stewart-mlergm}, and
\emph{igraph} \citep{Csardi-igraph}. In our simulation setting, we have three
network parameters of interest: $n$, or the number of nodes in the network;
$p_{\text{within}}$, or the probability of an edge between nodes in the same
communities; $p_{\text{between}}$, or the probability of an edge between nodes
in different communities. For our simulations, we specifically consider $n \in
\{20, 25, 50, 100, \ldots, 1000\}$ and both $p_{\text{within}},
p_{\text{between}} \in \{0.1, 0.15, 0.2, \ldots, 0.9\}$. Instead of considering
all combinations of parameter values, we select a value of each parameter as a
reference point ($n = 500$, $p_{\text{within}} = 0.75$, $p_{\text{between}} =
0.25$), measuring how estimator properties of interest (e.g., bias) change for
one parameter, while keeping the others fixed. We take the number of blocks and
the probability of community membership to be fixed at $k = 4$ and
$\frac{1}{k}$, respectively. (As suggested by our theory, we find that the
results of our approach are consistent for any fixed number of blocks, of
comparable sizes.)  Therefore, the latent community network (i.e., adjacency
matrix $A \in [0,1]^{n\times n}$ and community membership $\sigma \in [k]^n$)
for each simulation is drawn from the model space parameterized as
$\Theta\left(n,k,a,b,\beta\right)$, which satisfies the conditions described in
\citet{Gao-sbm-minimax_algo}\footnote{The GMZZ conditions also include a
parameter to control the differences across communities of the within- and
between-community connection probabilities. We omit this parameter as the
within- and between-community connection probabilities are both constant across
communities in our simulations. This restricted parameter space is discussed in
\citet{Gao-sbm-minimax_algo} as $\Theta_0$.}. More specifically, in our
simulations, $a \approx n\cdot p_{\text{within}}$, $b \approx n\cdot
p_{\text{between}}$, $k=4$, and $\Expect{\beta}=1$.

Given our network class and parameter set, we now define the data generation
process of interest that will describe the behavior of node level variables
across the network. We again consider the causal model defined in Figure
\ref{fig:graphical-model}, and the subsequent linear structural-equation model
defined in~\eqref{eqn:structural-equation}, which we restate for clarity:
\begin{equation*}
  Y_{i,t+1} = \alpha_0 + \alpha_1 Y_{i,t} + \beta\frac{\sum_{j}{\left(Y_{j,t}
  A_{ij}\right)}}{\sum_{j}{A_{ij}}} + \gamma_1^T C_i + \gamma_2^T X_{i} +
  \epsilon_{i,t+1}.
\end{equation*}
In each simulation, using~\citet{Sofrygin-simcausal}, we generate structural
equations with  parameters following a normal distribution $N(\mu, \sigma^2)$:
$\alpha_0 \sim N(1,1)$, $\alpha_1 \sim N(10,1)$, $\beta \sim N(0.1,1)$,
$\gamma_1 \sim N(10,1)$ and $\gamma_2 \sim N(1,1)$. Note that $C_i \in
\{1,\ldots,k\}$ starts as an integer (community identification) label, but for
the purpose of the regression is translated into a $k-1$ binary vector, and
therefore, $\gamma_1$ is also appropriately translated into a $k-1$ vector.
Additionally, we generate the following node-level variables:
$\epsilon_{i,t+1}\sim N(0,10)$ and $X_i \sim N(0,1)$, the latter which we treat
as a single variable capturing un-changing, network-irrelevant attributes for
each node. Finally, our goal is to estimate $\beta$, the coefficient for social
influence.

When the above is the structural-equation model generating our data, OLS will
provide an unbiased and consistent estimate of $\beta$, assuming we can observe
each of the variables relevant to the network (i.e., $A_{ij}$ and $C_i$) as well
as those that are irrelevant to the network but still relevant to behavior
($Y_{i,t}$, $Y_{j,t}$, and $X_{i}$). However, in practice, we do not observe
either $C_{i}$ or $X_{i}$, and therefore consider the OLS estimator of $\beta$
in \eqref{eqn:structural-equation} to be our ``Oracle'' estimator. Moreover, in
practice, we can obtain the OLS estimation of $\beta$ in
\eqref{eqn:effective-model}, which again we restate for clarity:
\begin{equation*}
  Y_{i,t+1} = \alpha_0 + \alpha_1 Y_{i,t} + \beta\frac{\sum_{j}{\left(Y_{j,t}
  A_{ij}\right)}}{\sum_{j}{A_{ij}}} + \gamma_0^T \hat{C}_i + \eta_{i,t+1} ~,
\end{equation*}
where $\hat{C}_i$ is an estimated location for node $i$ and the noise term
$\eta_{i,t+1}$ is now
\begin{equation*}
  \eta_{i,t+1} = \epsilon_{i,t+1} + \gamma_2^T X_{i} + (\gamma_1^T C_i -
  \gamma_0^T \hat{C}_i) ~.
\end{equation*}
It is the OLS estimator of $\beta$ in this equation that our proposed theory (in
conjunction with algorithms for deriving $\hat{C}_i$) provides sufficient
conditions for consistency and asymptotically unbiasedness; therefore we
consider this to be our ``Algorithm'' estimator. Critically, the bias present in
this estimator is induced by measurement error, resulting from the use of
$\hat{C}_i$ in place of the (unobserved) correct $C_i$; therefore we also
estimate
\begin{equation*}
  Y_{i,t+1} = \alpha_0 + \alpha_1 Y_{i,t} + \beta\frac{\sum_{j}{\left(Y_{j,t}
  A_{ij} \right)}}{\sum_{j}{A_{ij}}} + \gamma_1^T C_i + e_{i,t+1} ~,
\end{equation*}
and consider this our ``Correct'' estimator. Note that this estimator is the
limit of our consistent Algorithm estimator; additionally, unlike the Oracle
estimator, it is unable to condition on the (unobserved) $X_i$. Finally, we also
consider the OLS estimator of $\beta$ in
\begin{equation*}
  Y_{i,t+1} = \alpha_0 + \alpha_1 Y_{i,t} + \beta\frac{\sum_{j}{\left(Y_{j,t}
  A_{ij}\right)}}{\sum_{j}{A_{ij}}} + u_{i,t+1} ~,
\end{equation*}
which will have omitted variable bias because it incorrectly ignores the impact
of homophily all together; therefore, we consider this our ``Incorrect''
estimator.

Our primary goal in the simulations is to observe changes in the bias
experienced by each estimator (i.e., Oracle, Algorithm, Correct, and Incorrect)
described above, as conditions change. Additionally, we are also interested in
the relative variation of the estimators (as this has direct implications for
confidence intervals and coverage probabilities), and again how this variation
changes as conditions change. Finally, given that our Algorithm estimator trades
bias from omitted variables for that from measurement error, fundamentally its
efficacy will be related to its degree of (estimation) error in node locations;
therefore, we also are interested in observing how this estimation error changes
as conditions change.

\begin{figure}
  \centering
  \includegraphics[width=.99\linewidth]{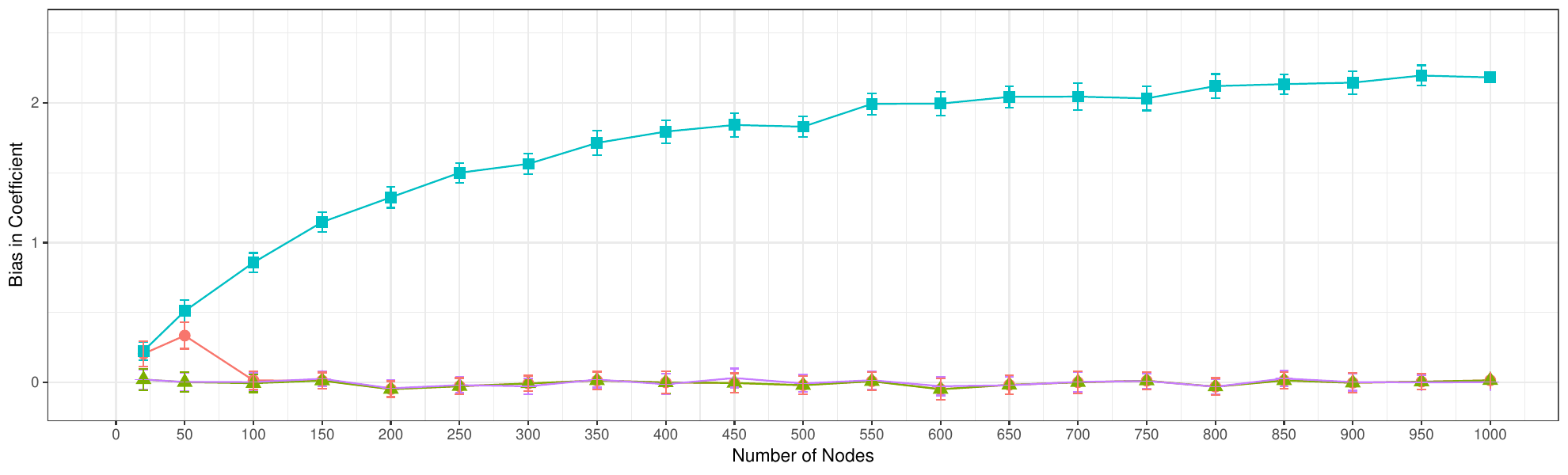}

  \includegraphics[width=.99\linewidth]{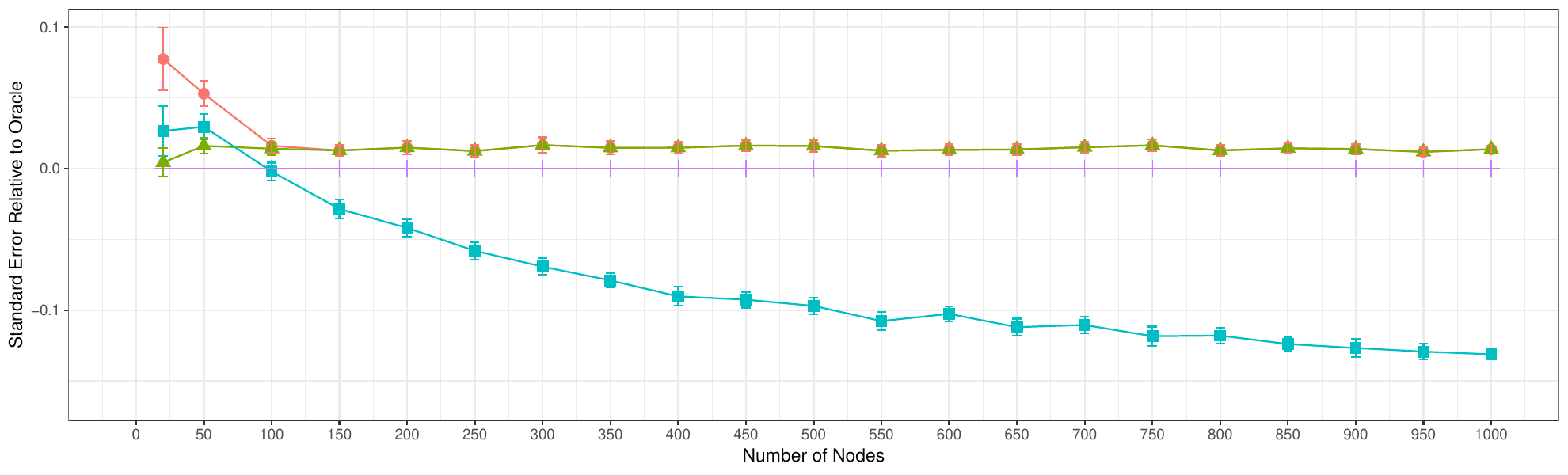}

  \includegraphics[width=.99\linewidth]{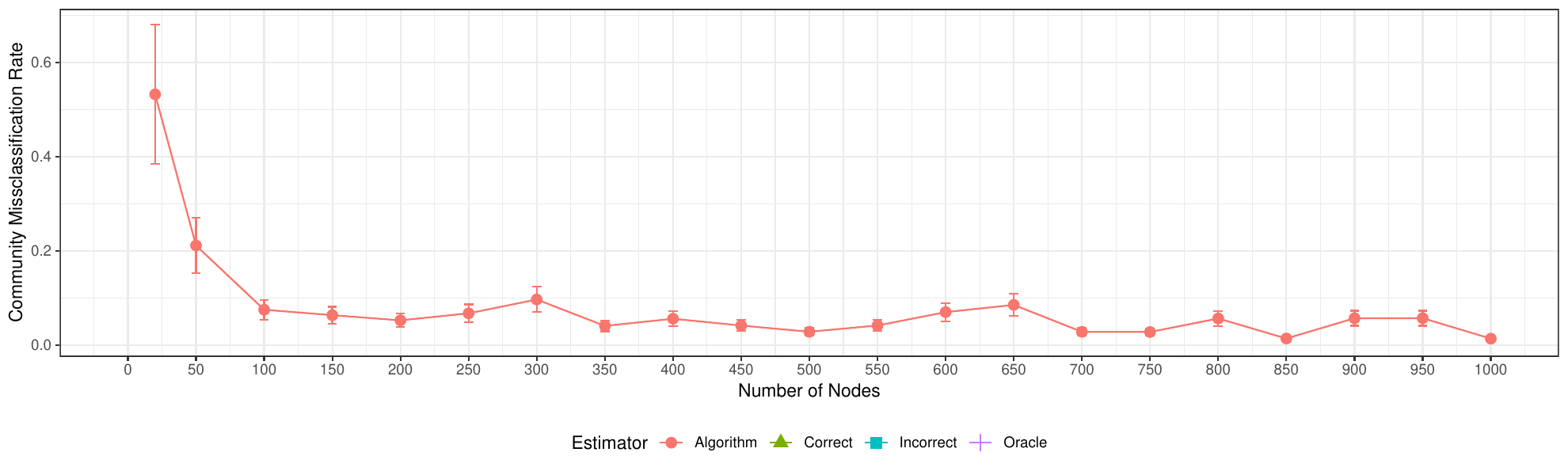}
  \caption{Comparison of the expected properties of the estimators where
  expectations is computed over $50$ random samples, allowing also for the
  formation of $95$\% confidence intervals. The parameter of interest is sample
  size $n$, which varies, while the latent community model parameters remain
  fixed ($k = 4, p_{\text{within}} = 0.75, p_{\text{between}} = 0.25$).}
  \label{fig:nodes}
\end{figure}

\subsection{Simulation Results}

In Figure~\ref{fig:nodes}, we observe how various outcomes of interest vary as
the sample size (number of nodes) increases (while the latent community model
parameters remain fixed at their reference values). We note that this simulation
complies with our assumed setting and therefore our predictions (based on our
theoretical results) of each estimator's behavior should be consistent with what
we observe. Let us begin by considering the first two plots, which show each
estimator's bias (top)---i.e., $\Expect{\hat{\beta}-\beta}$---and expected
standard error relative to that of the Oracle estimator (middle)---
$\Expect{\hat{\sigma}_{\hat{\beta}}-\hat{\sigma}_{\hat{\beta}_{Oracle}}}$--- as
a function of sample size. These plots confirm that the Oracle and Correct
estimators are unbiased at all sample sizes, but the Correct estimator has
larger variance, because it does not observe $X_i$. Additionally, the plots
confirm that ignoring the latent community (as in the Incorrect estimator) leads
to (omitted variable) bias at all sample sizes, which in our simulation amounts
to a bias exceeding $2$ numerical units for $\beta$ in the limit. (Since the
expected true value of $\beta$ across simulations, $\Expect{\beta} = 0.1$, a
bias of 2 units is a relative error of a remarkable $2,000\%$.) Beyond this
extreme bias, the estimator, additionally, becomes overconfident in its (biased)
estimation, which will lead to inaccurate confidence intervals and poor coverage
probability. Finally, the plots confirm that the Algorithm estimator (based on
estimated node locations) converges to the Correct estimator, and achieves
consistent and (asymptotically) unbiased estimation of peer-influence.

Figure~\ref{fig:nodes} also provides additional insight in the properties of the
Algorithm estimator, in comparing it to the other estimators. Importantly, we
observe that even at moderate sample sizes ($n = 100$) the estimator appears to
reach its asymptotic behavior (e.g., unbiasedness). Moreover, prior to reaching
this asymptotic behavior, the Algorithm and Incorrect estimators have similar
levels of bias, while the Algorithm estimator has larger variance. This implies
that the biases resulting from omitting the node locations and using estimated
locations (i.e. measurement error) are comparable, while the measurement error
induces larger variance; therefore, at small sample sizes, the Incorrect
estimator appears to provide a better estimation risk (with respect to loss in
mean squared error). However, as sample size increases, the trade-off between
these two sources of bias (and variance) begins to increasingly favor the
measurement error, and the Algorithm estimator provides better estimation risk.
We can see from the bottom plot in Figure~\ref{fig:nodes} that the risk of the
Algorithm estimator is, as expected, a function of the overall error in the node
locations. Additionally, this estimator reaches its asymptotic behavior
relatively quickly given the exponential decay in the measurement error.

\begin{figure}
  \centering
  \includegraphics[width=.99\linewidth]{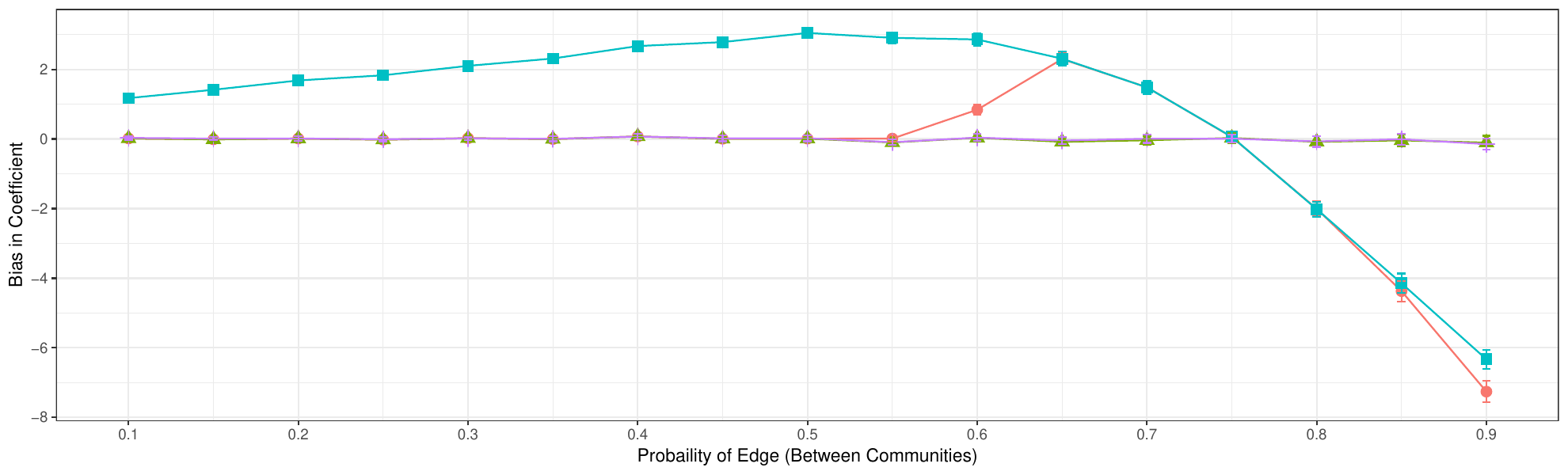}

  \includegraphics[width=.99\linewidth]{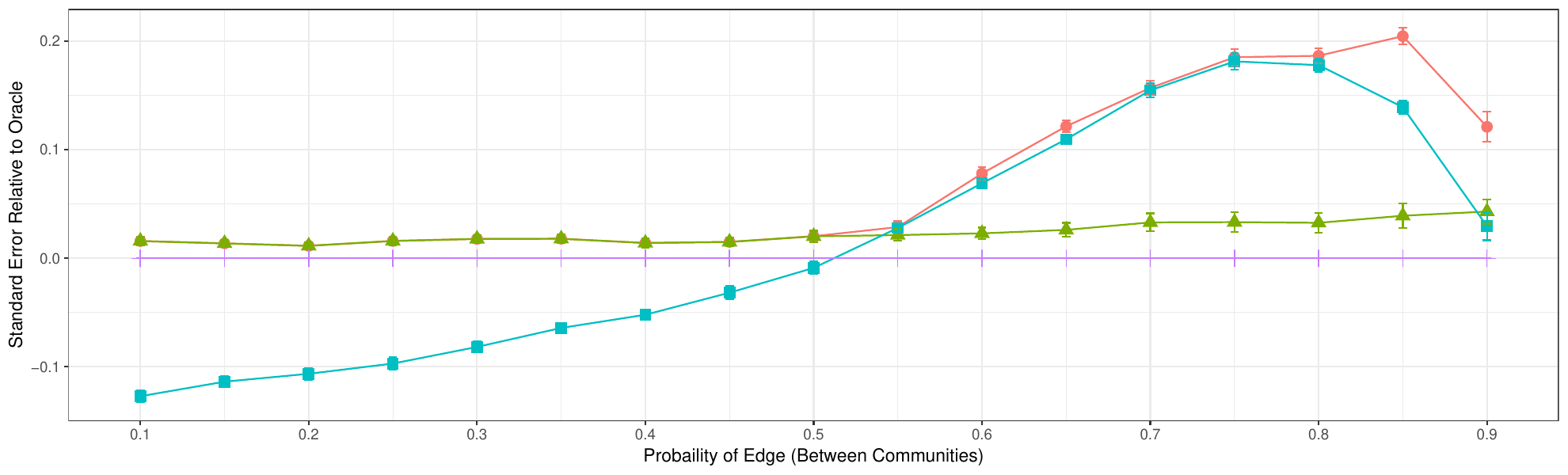}

  \includegraphics[width=.99\linewidth]{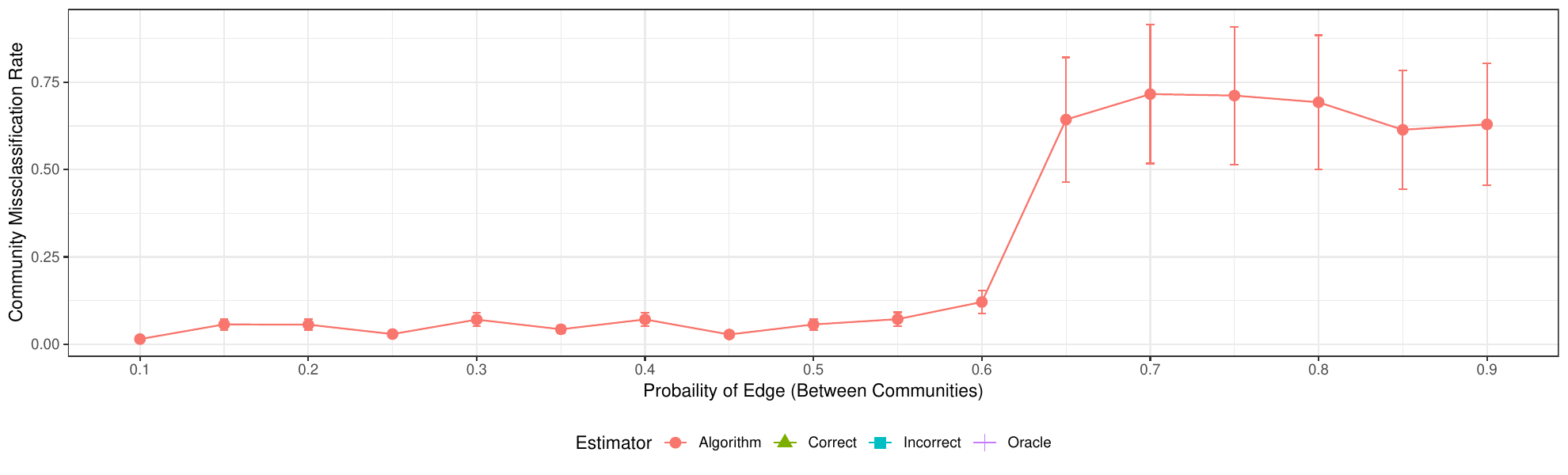}

  \caption{Comparison of the expected properties of the estimators where
  expectations is computed over $50$ random samples, allowing also for the
  formation of $95$\% confidence intervals. The parameter of interest is
  $p_{\text{between}}$, which varies, while the sample size and other latent
  community model parameters remain fixed ($n = 500, k = 4, p_{\text{within}} =
  0.75$).}
  \label{fig:bet}
\end{figure}

\paragraph{Assumption Violation} Although we are able to confirm our theoretical
guarantees when our (sufficient) conditions are met, we also aim to explore the
behavior of our estimator when these assumptions are violated. In Figure
\ref{fig:bet} we allow the probability of forming ties between nodes in
different communities ($p_{\text{between}}$) to vary, and we capture the same
three plots as before. As expected, when the between community ties
probabilities are low, the Algorithm estimator, as before, has behavior
equivalent to that of the Correct estimator. However, when the probability of
between community exceeds $0.5$, we notice that the Algorithm begins to exhibit
different behavior: both increased bias and variance. More specifically, we
notice it converges to the behavior of the Incorrect estimator, indicating that
the bias resulting from measurement error in the latent locations becomes as
large as that resulting from the omission of the locations. The bottom plot in
Figure~\ref{fig:bet} indicates that the Algorithm estimator's degradation in
behavior corresponds to its increase in latent location estimation error. The
source of this error, can be explained by revisiting~\eqref{eq:converg-rate-3}
as the bound it provides on the expected proportion of errors, includes the term
$I = D_{\frac{1}{2}}\left(\text{Ber}\left(\frac{a}{n}\right) \|
\text{Ber}\left(\frac{b}{n}\right) \right)$, where $\Expect{\frac{a}{n}} =
p_{\text{between}}$. More specifically, $I = \left(a-b\right)^2/(an)$ up to a
constant factor~\cite{Zhang-sbm_minimax}, therefore $I \rightarrow 0$ as
$p_{\text{between}} \rightarrow p_{\text{within}}$, increasing the probability
of location estimation errors.

Figure~\ref{fig:bet} also shows that when $p_{\text{between}} =
p_{\text{within}}$ (at $0.75$) the biases of the Incorrect and Algorithm
estimators are zero. At this point, the network is no longer homophilous (edges
within and between communities are equally likely), implying that there are no
longer arrows from $C_i$ and $C_j$ to $A_{ij}$ in the graphical causal model
(Figure~\ref{fig:graphical-model}). As a result there is no longer a confounding
backdoor pathway and there is no omitted variable bias. As $p_{\text{between}}$
increases beyond $p_{\text{within}}$, we see that the magnitude of the bias
begins to increase again, but in the opposite direction. This is because the
network is now increasingly heterophilous, and therefore $C_i$ and $C_j$ are
increasingly more negatively correlated. We observe that both the bias and
variance of the Algorithm estimator increases slightly beyond that of the
Incorrect estimator, which is likely because the Algorithm's assumption of
homophily is violated, and therefore it is grouping precisely the wrong nodes
together in a community. If there existed an approach that could achieve
consistent identification of latent communities for heterophilous networks,
consistent and (asymptotically) unbiased estimation of peer-influence can be
obtained with similar arguments to those in our theoretical results.

\begin{figure}
  \centering
  \includegraphics[width=.99\linewidth]{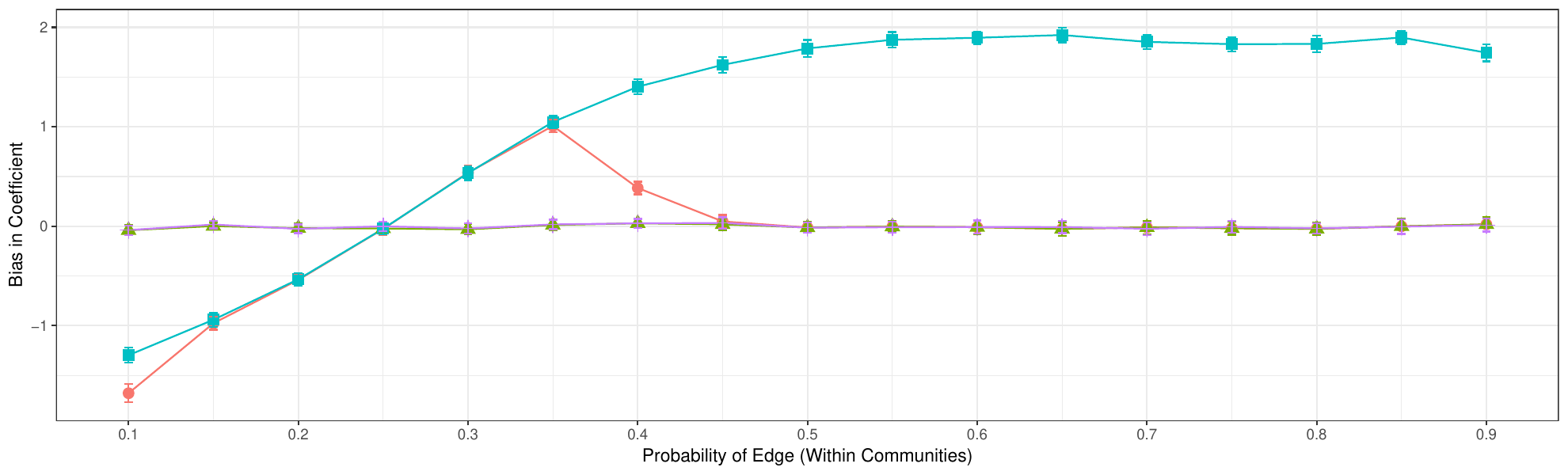}

  \includegraphics[width=.99\linewidth]{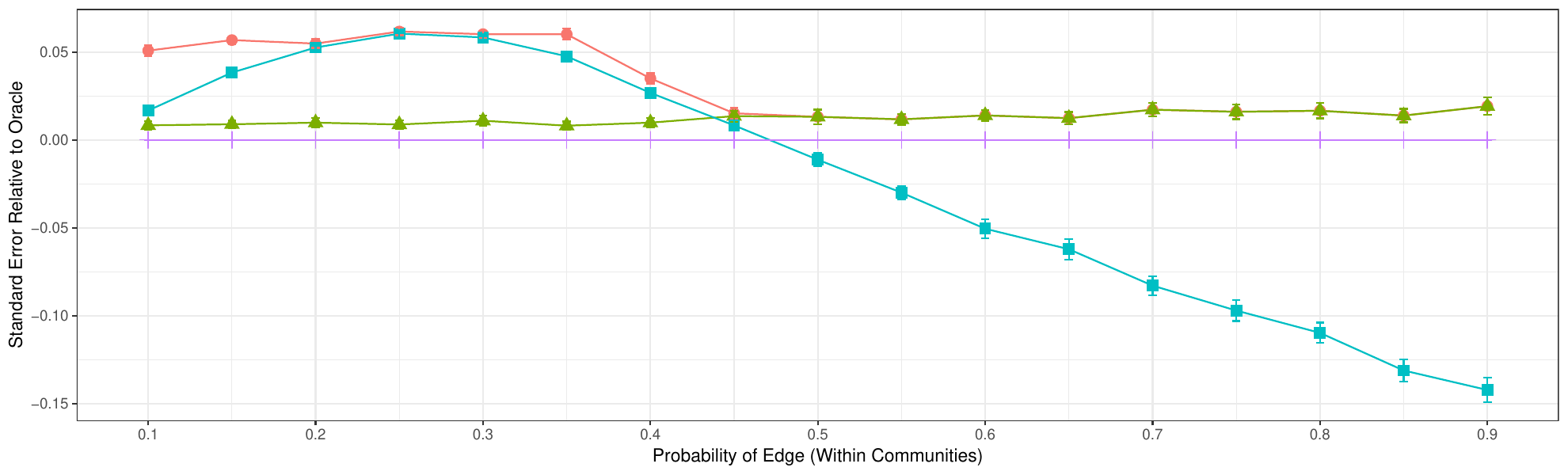}

  \includegraphics[width=.99\linewidth]{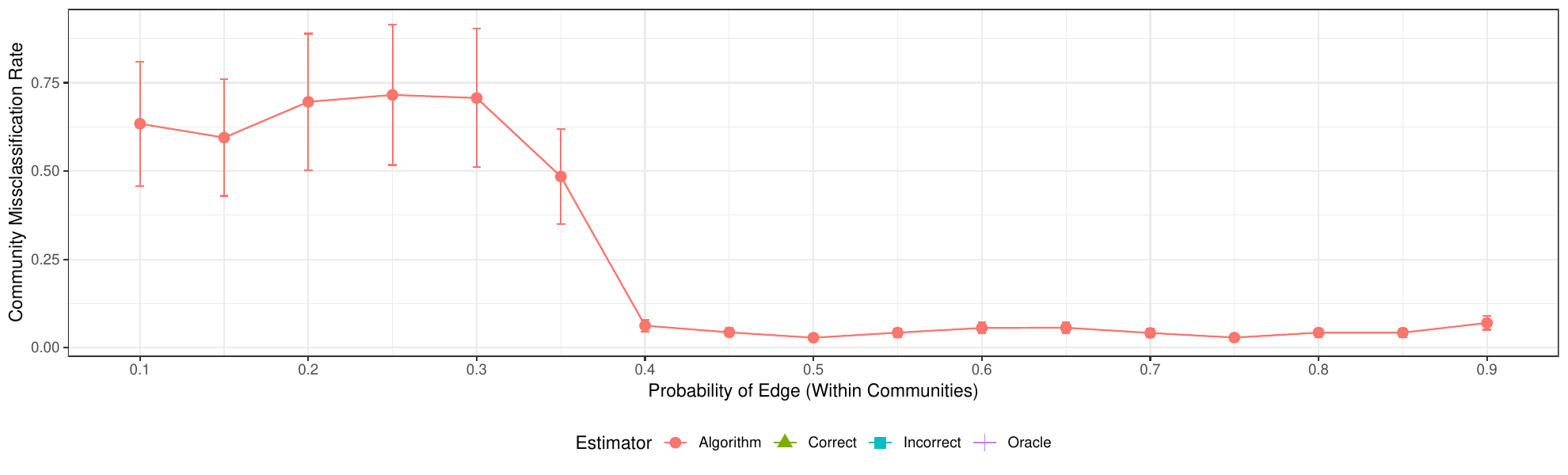}
  \caption{Comparison of the expected properties of the estimators where
  expectations is computed over $50$ random samples, allowing also for the
  formation of $95$\% confidence intervals. The parameter of interest is sample
  size $p_{\text{within}}$, which varies, while the sample size and other latent
  community model parameters remain fixed ($n = 500, k = 4, p_{\text{between}} =
  0.25$).}
  \label{fig:wit}
\end{figure}

In Figure~\ref{fig:wit} we allow the probability of forming ties between nodes
in the same community ($p_{\text{within}}$) to vary, which leads to conclusions
that are very similar to those for Figure~\ref{fig:bet} above, {\em mutatis
mutandis}. The additional insight that we obtain from Figure~\ref{fig:wit}, is
that the increased bias and variance in the Incorrect and Algorithm estimators
resulting from heterophily is smaller in magnitude than that in
Figure~\ref{fig:bet}. We suspect this is because, overall, the graph is more
sparse in the heterophilous facets of Figure~\ref{fig:wit} (as compared to those
in Figure~\ref{fig:bet}); therefore, there is less potential for peer-influence,
and the subsequent bias that results from its confoundment with
homophily.

\section{Discussion}
\label{sec:discussion}

We have shown that if a social network is generated by (a large class of) either
latent community models or continuous latent space models, and the pattern of
influence over that network then follows a linear model, it is possible to
obtain consistent and {\em asymptotically} unbiased estimates of the
social-influence parameter by controlling for estimates of the latent location
of each node.

These are, to our knowledge, the first theoretical results which establish
conditions under which social influence can be estimated from non-experimental
data without confounding, even in the presence of latent homophily. Previous
suggestions for providing such estimates by means of controlling for lagged
observations \citep{Valente-diffusion-of-innovations} or matching
\citep{aral2009distinguishing} are in fact all invalid in the presence of {\em
latent} homophily \citep{Homophily-contagion-confounded}. Instrumental variables
which are also associated with network location have been proposed
\citep{Tucker-network-externalities}; however, valid instruments are difficult
to obtain and even more difficult to verify, as fundamentally their satisfaction
of the exclusion restriction must be justified based on the specific context and
argued from (behavioral) theory. An alternative to full identification is to
provide {\em partial identification}
\citep{Manski-identification-for-prediction}, i.e., bounds on the range of the
social-influence coefficient.
\citet{VanderWeele-sensitivity-analysis-for-contagion} provides such bounds
under extremely strong parametric assumptions (among other things, $C_i$ must be
binary and it must not interact with anything);
\citet{Ver-Steeg-Galstyan-ruling-out-latent-homophily,
Ver-Steeg-Galstyan-tests-for-contagion} provide non-parametric bounds, but must
assume that each $Y_{i,t}$ evolves as a homogeneous Markov process, i.e., that
there is no aging in the behavior of interest. None of these limitations apply
to our approach.

Without meaning to diminish the value of our theoretical results, we feel it is
also important to be clear about their limitations. The following assumptions
were essential to our theoretical arguments:
\begin{enumerate}
\item The social network was generated {\em exactly} according to either a
  latent community model or a continuous latent space model.
\item We know whether it is a latent community model or a continuous latent
  space model.
\item We know either how many blocks there are (or how the number of blocks
  grows with $n$), or the latent space, its metric, and its link-probability
  function.
\item Fixed attributes of the nodes relevant to the behavior are either {\em
  fully} incorporated into the latent location, {\em or} stochastically
  independent of the location.
\item All of the relevant conditional expectation functions are linear.
\end{enumerate}

To augment our theory with empirical results, we also conduct a simulation study
specifically in the setting of networks generated according to a latent
community model. We find that if locations are estimated with a (deterministic)
minimax algorithm, our proposed estimator behaves as predicted by the theory,
when all assumptions are satisfied. However, we also find that the theory is not
fragile in the presence of small violations of the assumptions, e.g., the
(asymptotic) bias in the estimation increases smoothly as the network formation
process diverges from precisely a homophilous latent community. As a result, in
practice, even if the assumptions are not (perfectly) satisfied the estimates
should still exhibit bias reduction and (roughly) be ``close'' to the true
parameter of interest.

We suspect---though we have no proofs---that similar theoretical and empirical
results will hold for a somewhat wider class of well-behaved graphon network
models. (Graphon estimation is an active topic of current research
\citep{Choi-Wolfe-consistency-of-co-clustering,
Wolfe-Olhede-nonparametric-graphon-estimation}, but it has focused on estimating
the link-probability function $w$, rather than the latent locations $C$, though
see \citet{MEJN-Peixoto-generalized-communities} for a purely-heuristic
treatment.)  We also suspect such results will hold for nonlinear but smooth
conditional-expectation functions quite generally. (The simulations of
\citet{Worrall-homophily} indicate that the approach works with at least some
generalized linear models.)  Additionally, it's plausible that improved results
can be achieved with these well-behaved graphon models, when a subset of the
features relevant to tie formation (i.e., which impact node location) are
observed. We however note that incorporating these features will require
additional (careful) analysis, as any such feature may become redundant to
$\hat{C}$ and have an undesirable impact on the statistical properties of the
estimator. We also feel it is important to emphasize that there are many network
processes which are perfectly well-behaved, and are even very natural, which
fall outside the scope of our results; if, for instance, both ties $A_{ij}$ and
behaviors $Y_{i,t}$ are influenced by a latent variable $C_i$ which has {\em
both} continuous and discrete coordinates, there is no currently known way to
consistently estimate the whole of $C_i$.

Despite these disclaimers, we wish to close by emphasizing the following point.
In general, the strength of social influence cannot be estimated from
observational social network data, because any feasible distribution over the
observables can be achieved in infinitely many ways that trade off influence
against latent homophily. What we have shown above is that {\em if} the network
forms according to either of two standard models, and the rest of our
assumptions hold, this result can be evaded, because the network itself makes
all the relevant parts of the latent homophilous attributes manifest. To the
best of our knowledge, this is the {\em first} situation in which the strength
of social influence can be consistently estimated in the face of latent
homophily---the first, but we hope not the last.

\section{Proofs}
\label{sec:proofs}

\noerror*
\begin{proof}
We are chiefly concerned with $\hat{\beta}_{OLS}$, the ordinary least squares
estimate of $\beta$ in
\begin{align*}
  Y_{i,t+1} &= \alpha_0 + \alpha_1 Y_{i,t} + \beta\frac{\sum_{j}{\left(Y_{j,t}
  A_{ij}\right)}}{\sum_{j}{A_{ij}}} + \gamma_0^T \hat{C}_i +
  \overbrace{\epsilon_{i,t+1} + \gamma_2^T X_{i} + \left(\gamma_1^T C_i -
  \gamma_0^T \hat{C}_i\right)}^{\eta_{i,t+1} } \\
  &= \alpha_0 + \alpha_1 Y_{i,t} + \beta\frac{\sum_{j}{\left(Y_{j,t}
  A_{ij}\right)}}{\sum_{j}{A_{ij}}} + \gamma_1^T \hat{C}_i +
  \overbrace{\epsilon_{i,t+1} + \gamma_2^T X_{i} - \gamma_1^T\left( \hat{C}_i -
  C_i \right)}^{\eta_{i,t+1}},
\end{align*}
where $\hat{C}_i$ is an estimated location for node $i$, $\eta_{i,t+1}$ is the
(unobserved) noise term, and the equality follows from recognizing that $C =
\hat{C}_i - \left( \hat{C}_i - C_i \right)$. By the assumption that $\Prob{C
\neq \hat{C}} = 0$, allowing for the replacement of $\hat{C}$ with $C$, this
becomes
\begin{equation*}
  Y_{i,t+1} = \alpha_0 + \alpha_1 Y_{i,t} + \beta\frac{\sum_{j}{\left(Y_{j,t}
  A_{ij}\right)}}{\sum_{j}{A_{ij}}} + \gamma_1^T C_i +
  \overbrace{\epsilon_{i,t+1} + \gamma_2^T X_{i}}^{\eta_{i,t+1}}.
\end{equation*}
Given that $X_i \indep A_{ij}, Y_{j,t} | C_i,C_j$, we have
$\Cov{\frac{\sum_{j}{\left(Y_{j,t}A_{ij}\right)}}{\sum_{j}{A_{ij}}},
\eta_{i,t+1}} = 0$ and therefore the OLS estimator for $\beta$ is unbiased and
consistent.
\end{proof}

\altrsanderrors*
\begin{proof}
First, recognize that
\begin{align}
  &\Cov{\frac{\sum_{j}{\left(Y_{j,t}A_{ij}\right)}} {\sum_{j}{A_{ij}}},
  (\gamma_1^T C_i - \gamma_0^T \hat{C}_i) \middle| A, Y_{i,t}} \nonumber \\
  &~~~=\frac{\sum_{j}{A_{ij}\Cov{Y_{j,t}, (\gamma_1^T C_i - \gamma_0^T
  \hat{C}_i) \middle| A, Y_{i,t}}}}{\sum_{j}{A_{ij}}},
  \label{eqn:omitted-variable-bias-in-terms-of-covariances}
\end{align}
which follows from the linearity of covariance and the fact that $A$ is
conditioned on (and therefore constant). Therefore, we consider the terms in the
sum in the numerator:
\begin{eqnarray}
  \lefteqn{
    \nonumber \Cov{Y_{j,t}, (\gamma_1^T C_i - \gamma_0^T \hat{C}_i)|A, Y_{i,t}}}
    & & \\
    & = & \Cov{Y_{j,t}, \gamma_1^T C_i |A, Y_{i,t}
  } \label{eqn:dropping-chat-from-cond-cov}\\
  & = & \gamma_1^T \Cov{Y_{j,t},  C_i |A, Y_{i,t}}
  \label{eqn:pulling-gamma-1-out-of-cond-cov}
\end{eqnarray}
where~\eqref{eqn:dropping-chat-from-cond-cov} follows because we are
conditioning on $A$ ($\hat{C}_i$ is a deterministic function of $A$) and
additive constants do not change covariances.
Additionally,~\eqref{eqn:pulling-gamma-1-out-of-cond-cov} follows by linearity
of covariance.

We are thus interested in the conditional covariance between $Y_{j,t}$ and
$C_i$. We can at this point use the fact that \eqref{eqn:structural-equation} is
a {\em linear} structural equation system. This allows us to use the Wright
rules \citep{Wright-path-coefficients} to ``read off'' (conditional) covariances
from the DAG corresponding to the structural equations
\citep{Moran-on-path-coefficients}. Briefly stated, to find the covariance
between two variables $F$ and $G$ conditional on a set of variables $H$, these
rules require us to (i) find all paths between $F$ and $G$ in the DAG, (ii)
discard those paths which are ``closed'' when conditioning on $H$, (iii)
multiply the linear regression coefficients encountered at each step along a
path, (iv) multiply by a ``source'' variance for the common ancestor of all the
variables along a path (conditional on $H$), when one exists, {\em or} (v)
multiply by the conditional covariance of two ``sources'' linked by conditioning
on a collider, and (vi) sum up over paths. (For the notion of a path in a DAG
being ``open'' or ``closed'' when conditioning on a set of variables, see, e.g.,
\citet[Definition 1, p.\ 106]{Pearl-on-causal-inference}.)  Before presenting
the relevant paths, it is convenient to introduce the abbreviation $d_j =
\sum_{i}{A_{ij}}$ for the ``degree'' of node $j$, i.e., the number of social
ties it has.

\begin{itemize}
\item {\em Path}: $Y_{j,t} \leftarrow C_j \rightarrow A_{ij} \leftarrow C_i$.
  {\em Contribution}: $ \Cov{C_i, C_j|A} \gamma_1$.

\item {\em Path}: $Y_{j,t} \leftarrow Y_{j, t-1} \leftarrow Y_{i, t-2}
  \leftarrow C_i$. {\em Contribution}: $\alpha_1 \frac{\beta A_{ij}}{d_j}
  \Var{C_i|A} \gamma_1$.

\item {\em Path}: $Y_{j,t} \leftarrow Y_{j, t-1} \leftarrow Y_{j, t-2}
  \leftarrow Y_{i, t-3} \leftarrow C_i$. {\em Contribution}: $\alpha_1^2
  \frac{\beta A_{ij}}{d_j}  \Var{C_i|A} \gamma_1$.

\item {\em Path}: $Y_{j,t} \leftarrow Y_{j, t-1} \leftarrow \ldots \leftarrow
  Y_{j, t-h} \leftarrow Y_{i, t-h-1} \leftarrow C_i$. {\em Contribution}:
  $\alpha_1^h \frac{\beta A_{ij}}{d_j}  \Var{C_i|A} \gamma_1$.

\item {\em Path}: $Y_{j,t} \leftarrow Y_{i, t-2} \leftarrow \ldots \leftarrow
  Y_{i, t-h} \leftarrow C_i$. {\em Contribution}: $\frac{\beta A_{ij}}{d_j}
  \alpha_1^{h-2}\Var{C_i|A} \gamma_1$.

\item {\em Path}: $Y_{j,t} \leftarrow Y_{l, t-1} \leftarrow C_l \rightarrow
  A_{li} \leftarrow C_i$. {\em Contribution}: $\frac{\beta A_{jl}}{d_j}\Cov{C_l,
  C_i|A} \gamma_1$. (This must be summed over all possible nodes $l$.)

\item {\em Path}: $Y_{j,t} \leftarrow Y_{j, t-1} \leftarrow Y_{l, t-2}
  \leftarrow C_l \rightarrow A_{li} \leftarrow C_i$. {\em Contribution}:
  $\alpha_1\frac{\beta A_{jl}}{d_j}\Cov{C_l, C_i|A} \gamma_1$. (Similar paths
  extending back into the past add powers of $\alpha_1^2, \alpha_1^3$, etc. This
  must also be summed over all possible nodes $l$.)

\item {\em Path}: $Y_{j,t} \leftarrow Y_{l, t-1} \leftarrow Y_{l, t-2}
  \leftarrow C_l \rightarrow A_{li} \leftarrow C_i$. {\em Contribution}:
  $\alpha_1 \frac{\beta A_{jl}}{d_j}\Cov{C_l, C_i|A}  \gamma_1$. (Similar paths
  extending back into the past add powers of $\alpha_1^2, \alpha_1^3$, etc.)
\end{itemize}

From this enumeration, two things are clear: $1)$ {\em all} the paths lead to
terms involve a single power of $\gamma_1$, and $2)$ every term involves a
factor of either $\Cov{C_j, C_i|A}$ or $\Var{C_i|A}$. Combining paths with the
same source terms, we therefore have
\begin{eqnarray*}
  \lefteqn{
    \Cov{Y_{j,t},  C_i |A, Y_{i,t}}
  } & &\\
  & = &  \Cov{C_i, C_j|A } \gamma_1\\
  \nonumber & & + \left(\sum_{h=1}^{T-1}{\alpha_1^h \frac{\beta A_{ij}}{d_j}}+
  \sum_{h=3}^{T-1}{\frac{\beta A_{ij}}{d_j} \alpha_1^{h-2}}\right)
  \Var{C_i|A}\gamma_1\\
  \nonumber & & + \sum_{l \neq i, j}{\left(\sum_{h=0}^{T-1}{\alpha_1^h
  (1+\alpha_1) \frac{\beta A_{jl}}{d_j}} \right) \Cov{C_l, C_i|A}} \gamma_1\\
  & = & \Cov{C_i, C_j|A} \gamma_1\\
  \nonumber && + \xi_{ij} \Var{C_i|A} \gamma_1\\
  \nonumber && + \sum_{l \neq i, j}{\zeta_{ijl} \Cov{C_l, C_i|A}} \gamma_1
\end{eqnarray*}
introducing $\xi_{ij}$ and $\zeta_{ijl}$ as the abbreviations for the
appropriate sums. Substituting back into
\eqref{eqn:pulling-gamma-1-out-of-cond-cov} amounts to multiplying every term
here by $\gamma_1^T$ from the left. Substituting in turn into
\eqref{eqn:omitted-variable-bias-in-terms-of-covariances} yields the promised
lemma.
\end{proof}

{\em Remark:} The form of the covariance $\Cov{Y_{j,t}, C_i|A, Y_{i,t}}$ is
somewhat complicated, because it turns out that many paths connect $Y_{j,t}$ and
$C_i$. Most of these paths would, however, be closed if we also conditioned on
$Y_{j, t-1}$ and $Y_{i, t-1}$. Conditioning on lagged values of $Y$ for both ego
and alters in this way is sometimes done by practitioners, and would indeed
leave open only the path $Y_{j, t} \leftarrow C_j \rightarrow A_{ij} \leftarrow
C_i$. This would simplify the conditional covariance between $Y_{j,t}$ and $C_i$
to just $\gamma_1^T \Cov{C_i, C_j|A} \gamma_1$. However, conditioning on these
lagged values would mean altering the regression specification, and with it the
coefficients and their interpretation. In particular, if autoregressive effects
within nodes are strong, then $Y(j,t)$ and $Y(j, t-1)$ will be strongly
correlated, which will introduce its own potential biases into the estimation of
$\beta$. The net result {\em may} be to reduce the bias, but this would require
detailed calculation. Since (as we show below) we are able to get consistent
estimation of $\beta$ without introducing these lagged terms, we do not pursue
this further here.

\asycovblocks*
\begin{proof}
First, we let $M_{n} = \sum_i^n{\mathbbm{1}\{\hat{C}_i \ne C_i\}}$, then from
\eqref{eq:converg-rate-2}--\eqref{eq:converg-rate-3}, we have
\begin{equation*}
  \Expect{M_{n}/n} \leq e^{-cn}
\end{equation*}
for an appropriate constant $c > 0$ (and large enough $n$), which implies
\begin{equation*}
  \Expect{M_{n}} \leq n e^{-cn}.
\end{equation*}
We now turn our focus to the probability that $M \geq 1$:
\begin{equation*}
  \label{eqn:probability-of-any-errors-is-exp-small}
  \begin{split}
    \Prob{M_{n}\geq 1}&\leq \Expect{M_{n}}/1 \quad(\text{Markov's Inequality})\\
    &\leq n e^{-cn}  \\
    &= e^{-cn+\log{n}} \\
    &= e^{-O(n)}.
  \end{split}
\end{equation*}

Therefore, the probability of making any latent location estimation errors at
all goes to zero exponentially fast in $n$, and we note that it does so almost
surely. Indeed, the almost sure convergence follows since $\sum_{n}{n e^{-cn}}$
is finite\footnote{To see this, differentiate the geometric series
$\sum_{n}{e^{-cn}}$ with respect to $c$.}, and the Borel-Cantelli lemma
\citep[Theorem 7.3.10a, p.\ 288]{Grimmett-Stirzaker} tells us that with
probability $1$, $M_n \geq 1$ only finitely often, i.e., that $M_n \rightarrow
0$ almost surely. Therefore, with probability tending to one almost surely, as
$n\rightarrow\infty$, $\hat{C} = C$. As a direct consequence, $\Cov{C_i, C_j|
\hat{C}_i, \hat{C}_j} \xrightarrow{a.s.} 0$. We note that although we have
almost sure convergence in
Lemma~\ref{lemma:asymptotic-covariance-for-block-models}, only weaker
consistency (convergence in probability) is required for the results that build
atop this Lemma.
\end{proof}

\fincovblocks*
\begin{proof}
The second part of the lemma follows automatically from the first part, and
the fact that assuming the GMZZ conditions means that the requirements of
Lemma \ref{lemma:asymptotic-covariance-for-block-models} are satisfied,
implying that $\delta(n) = e^{-O(n)}$. Accordingly, we focus on establishing
the first part of the lemma.

We now evoke the law of total covariance and decompose
\begin{eqnarray}
  \nonumber \lefteqn{\Cov{C_i, C_j|A}} & & \\
  & = & \Expect{\Cov{C_i, C_j|A,G_n}|A} + \Cov{\Expect{C_i|A,G_n},
  \Expect{C_j|A,G_n}|A}, \label{eqn:law-of-total-covariance}
\end{eqnarray}
where $G_n=1$ if all the nodes are assigned to their correct blocks (so $C_i =
\hat{C}_i$ for all $i$) and $G_n=0$ otherwise. Given this decomposition, we will
need to make a series of steps, dealing in turn with the expected covariance and
the covariance of the expectations.

\underline{Step 1:}
Looking at the conditional covariance, we know
\begin{eqnarray*}
  \lefteqn{\Expect{\Cov{C_i, C_j|A, G_n}|A}} & &\\
  \nonumber & =  & \Prob{G_n=1|A}\Cov{C_i, C_j|A, G_n=1} + \Prob{G_n=0|A}
  \Cov{C_i, C_j|A, G_n=0}.
\end{eqnarray*}
We also recognize that
\begin{equation*}
  \Cov{C_i, C_j|A, G_n=1} = \Cov{\hat{C}_i, \hat{C}_j|A, G_n=1} = 0,
\end{equation*}
where the first equality follows from $G_n=1$ (i.e., $\hat{C}_i = C_i~\forall
i$) and the second equality follows because $\hat{C}_i$ and $\hat{C}_j$ are
functions $A$, which we condition on. Next we note that
\begin{equation*}
  \Cov{C_i, C_j|A, G_n=0} \neq 0;
\end{equation*}
however, because $C_i$ and $C_j$ are ``dummy'' or indicator vectors, they are
points on the corners of the $k-1$ dimensional simplex (or the origin).
Moreover, $\Cov{C_i, C_j|A, G_n=0}$ is a $k \times k$ covariance matrix, whose
entries are bounded above by $1$ and below by $-1$. Therefore, the magnitude of
$\lVert \Cov{C_i, C_j|A, G_n=0} \rVert$ is bounded by a constant (with respect
to $n$) whose value depends on the specific norm $\lVert \cdot \rVert$ used to
measure magnitude. Therefore, combining the results for $G_n=1$ and $G_n=0$, we
have
\begin{equation}
  \Expect{\Cov{C_i, C_j|A, G_n}|A} = 0 + O(\delta(n)).
  \label{eqn:expected-conditional-covariance-is-order-of-error-probability}
\end{equation}

\underline{Step 2:}
Turning to the conditional expectations, we similarly know that
\begin{equation}
  \Expect{C_i|A, G_n=1} = \hat{C}_i, \label{eqn:location-given-truth}
\end{equation}
because when $G_n=1$, $\hat{C}_i = C_i~\forall i$. We can also {\em define} a
new variable $\tilde{C}_i$ such that
\begin{equation}
  \label{eqn:location-given-not-truth}
  \tilde{C}_i \equiv \Expect{C_i|A, G_n=0}.
\end{equation}
This new random variable $\tilde{C}_i$ is a function of $A$, and takes values
within the interior of the convex hull of the  $k-1$ dimensional simplex and the
origin (rather than at the simplex's corners and the origin). Because $G_n$ is
an indicator variable, we can combine \eqref{eqn:location-given-truth} and
\eqref{eqn:location-given-not-truth} to write
\begin{equation}
  \Expect{C_i|A, G_n} = \hat{C}_i G_n + (1-G_n) \tilde{C}_i
  \label{eqn:representation-in-gn}
\end{equation}
and similarly for $\Expect{C_j|A, G_n}$. Using \eqref{eqn:representation-in-gn}
we can compute the covariance between the conditional expectations of the node
locations:
\begin{eqnarray}
  \nonumber  \lefteqn{\Cov{\Expect{C_i|A, G_n}, \Expect{C_j|A, G_n}|A}} & &\\
  \nonumber & = & \Cov{\hat{C}_i G_n + \tilde{C}_i(1-G_n), \hat{C}_j G_n +
  \tilde{C}_j (1-G_n)|A}\\
  \nonumber  & = & \Cov{\hat{C}_i G_n, \hat{C}_j G_n|A} + \Cov{\tilde{C}_i
  (1-G_n), \tilde{C}_j (1-G_n)|A}\\
  \nonumber & & + \Cov{\hat{C}_i G_n, \tilde{C}_j (1-G_n)|A} + \Cov{\tilde{C}_i
  (1-G_n), \hat{C}_j G_n|A}\\
  & = & \hat{C}_i\Var{G_n|A} \hat{C}_j^T + \tilde{C}_i \Var{1-G_n|A}
  \tilde{C}_j^T \label{eqn:pull-out-the-vectors}\\
  \nonumber & & + \hat{C}_i\Cov{G_n, 1-G_n|A} \tilde{C}_j^T + \tilde{C}_i
  \Cov{1-G_n, G_n|A} \hat{C}_j^T \\
  & = & O \left(\Var{G_n|A}\right) + O\left(\Var{1-G_n|A}\right)
  \label{eqn:simplex-has-bounded-magnitude}\\
  \nonumber & & + O\left(\Cov{G_n, 1-G_n|A}\right)+ O\left(\Cov{1-G_n,
  G_n|A}\right)\\
  & = & O(\delta(n)).
  \label{eqn:variances-and-covariances-are-order-of-the-error-probability}
\end{eqnarray}
\eqref{eqn:pull-out-the-vectors} follows from the fact that the four vectors
---$\hat{C}_i, \tilde{C}_i, \hat{C}_j$ and $\tilde{C}_j$---are all functions of
$A$ and therefore conditionally constant. Moreover,
\eqref{eqn:simplex-has-bounded-magnitude} follows from the fact that these
vectors all lie within the convex hull of the $k-1$ dimensional simplex and the
origin, and therefore their outer products---$\hat{C}_i\hat{C}_j^T$,
$\tilde{C}_i\tilde{C}_j^T$, $ \hat{C}_i\tilde{C}_j^T$ and $\tilde{C}_i
\hat{C}_j^T$---are also bounded by a constant (with respect to $n$). Finally,
\eqref{eqn:variances-and-covariances-are-order-of-the-error-probability} follows
from two realizations. First, that $1-G_n$ is a binary variable whose
expectation is $O(\delta(n))$, so $\Var{1-G_n|A}=\Var{G_n|A} = \delta(n)
(1-\delta(n)) = O(\delta(n))$. Secondly, since $G_n(1-G_n)=0$ always,
$\Cov{G_n,1-G_n|A} = \Expect{G_n(1-G)_n|A} - \Expect{G_n|A}\Expect{1-G_n|A} = -
(1-\delta(n))\delta(n) = O(\delta(n))$.

Thus plugging
\eqref{eqn:expected-conditional-covariance-is-order-of-error-probability} and
\eqref{eqn:variances-and-covariances-are-order-of-the-error-probability} into
\eqref{eqn:law-of-total-covariance}, we have
\begin{equation*}
  \Cov{C_i, C_j|A} = O(\delta(n)) + O(\delta(n)) = O(\delta(n))
\end{equation*}
\end{proof}

\begin{lemma}
  \label{lemma:label-variance-for-block-models} Suppose that the assumptions
  from Section \ref{sec:setting} hold, the network forms according to a latent
  community model, and $\hat{C}$ can be estimated with error rate $\delta(n)$.
  Then $\Var{C_i|A} = O(\delta(n))$. If the latent community model also
  satisfies the GMZZ conditions and a minimax algorithm is used to estimate
  $\hat{C}$, then $\Var{C_i|A} = e^{-O(n)}$.
\end{lemma}
\begin{proof}
  The proof runs along the same lines as that of Lemma
  \ref{lemma:covariance-for-block-models}, albeit with somewhat less algebra,
  and so only sketched. We can write $\Var{C_i|A} = \Expect{\Var{C_i|A,G_n}|A} +
  \Var{\Expect{C_i|A,G_n}|A}$. $\Var{C_i|A,G_n=1} = 0$, because, conditional on
  $G_n=1$, $C_i = \hat{C_i}$ which is a function of $A$. If $G_n=0$, however,
  the variance of $C_i$ is bounded, since every possible value of $C_i$ is a
  corner on the simplex (or the origin), hence $\Expect{\Var{C_i|A,G_n}|A} =
  O(\delta(n))$. Similarly, $\Expect{C_i|A,G_n=1} = \hat{C}_i$, which is
  constant (conditional on $A$) and does not contribute to the
  conditional-on-$A$ variance, while $\Expect{C_i|A,G_n=0}$, which is random
  with respect to $A$, is still bounded within the convex hull of the simplex
  and the origin. Thus $\Var{C_i|A} = O(\delta(n))$ over-all. Further assuming
  the GMZZ conditions tells us $\delta(n) = e^{-O(n)}$.
\end{proof}

\thmblocks*
\begin{proof}
As in Lemma~\ref{lemma:no-error-no-problem}, we are again chiefly concerned with
$\hat{\beta}_{OLS}$, the ordinary least squares estimate of $\beta$ in
\begin{equation}
  \label{eqn:theorem-blocks-estimatable}
  Y_{i,t+1} = \alpha_0 + \alpha_1 Y_{i,t} + \beta\frac{\sum_{j}{\left(Y_{j,t}
  A_{ij}\right)}}{\sum_{j}{A_{ij}}} + \gamma_0^T \hat{C}_i +
  \overbrace{\epsilon_{i,t+1} + \gamma_2^T X_{i} + \left(\gamma_1^T C_i -
  \gamma_0^T \hat{C}_i\right)}^{\eta_{i,t+1} }
\end{equation}
where $\hat{C}_i$ is an estimated location for node $i$, $\eta_{i,t+1}$ is the
(unobserved) noise term. Moreover, we know that
\begin{align}
  \Expect{\hat{\beta}_{OLS} \middle| A}  &= \beta + O\left(\Cov{\frac{\sum_{j}
  {\left(Y_{j,t}A_{ij}\right)}}{\sum_{j}{A_{ij}}}, \eta_{i,t+1}}\middle|A
  \right) \label{eqn:theorem-blocks-ols-estimate}\\
  &= \beta + O\left(\Cov{\frac{\sum_{j}{\left(Y_{j,t}A_{ij}\right)}}{\sum_{j}
  {A_{ij}}}, \left(\gamma_1^T C_i - \gamma_0^T \hat{C}_i\right)}\middle|A
  \right) \label{eqn:theorem-blocks-setting}\\
  &= \beta + O\left(\gamma_1^T \Cov{C_i, C_j|A} \gamma_1\right) + O\left(
  \gamma_1^T \Var{C_i|A} \gamma_1\right) + O\left(\gamma_1^T \Cov{C_i, C_l|A}
  \gamma_1\right)\label{eqn:theorem-blocks-lemma-altrsanderrors}\\
  &= \beta + O\left(\delta(n)\right)
  \label{eqn:theorem-blocks-lemma-finite-covariance}
\end{align}
where \eqref{eqn:theorem-blocks-ols-estimate} follows from the definition of the
OLS estimate for $\beta$ in \eqref{eqn:theorem-blocks-estimatable},
\eqref{eqn:theorem-blocks-setting} follows from the assumptions of the setting
(chiefly \eqref{eqn:screening-egos-attributes-from-alters-behavior-2}),
\eqref{eqn:theorem-blocks-lemma-altrsanderrors} follows from Lemma
\ref{lemma:covariance-between-altrs-behavior-and-estimation-errors}, and finally
\eqref{eqn:theorem-blocks-lemma-finite-covariance} follows from Lemmas
\ref{lemma:covariance-for-block-models} and
\ref{lemma:label-variance-for-block-models}. Moreover we have that
$O\left(\delta(n)\right)$ can be made exponentially small, and in only a
polynomial cost in computational time, (\S \ref{sec:communities} above).
Therefore, the bias in $\hat{\beta}_{OLS}$ is itself exponentially small in $n$.
Hence $\hat{\beta}_{OLS}$ will be asymptotically unbiased and consistent as
$n\rightarrow\infty$.
\end{proof}

\thmspace*
\begin{proof}
As in the proof of Theorem \ref{theorem:asymptotic-consistent-for-block-models},
it will be enough to show that both $\Cov{C_i, C_j|A} \rightarrow 0$ and
$\Var{C_i|A} \rightarrow 0$. To do so, we showed that $\Cov{C_i, C_j|A}$ and
$\Var{C_i|A}$ were both $O(\delta(n))$, where $\delta(n)$ was the probability of
community discovery mis-labeling any nodes at all. We cannot expect such exact
recovery of the latent variables in a continuous model, so we will work instead
with $\delta(\epsilon_n, n)$, the probability that all estimated positions are
within $\epsilon_n$ of the true positions, and let $\epsilon_n \rightarrow 0$ at
a suitable rate.

To be specific, define $\delta(n, \epsilon)$ as $\Prob{\max_{i \in 1:n}{\|C_i -
\hat{C}_i\| \geq \epsilon}}$, where $\hat{C}_i$ is the maximum likelihood
estimate of $C_i$. By \eqref{eqn:asta-bound}
\begin{equation*}
  \delta(n, \epsilon) \leq \mathcal{N}(n, \epsilon) e^{-\kappa \epsilon n^2}
\end{equation*}
where $\mathcal{N}(n, \epsilon)$ is polynomial in both $n$ and in $1/\epsilon$.
Now fix a sequence $\epsilon_n > 0$ such that $\epsilon_n \rightarrow 0$ as
$n\rightarrow \infty$, while $\epsilon_n n^2 \rightarrow \infty$ at least
polynomially fast in $n$. (For instance, but not necessarily optimally,
$\epsilon_n = n^{-1}$.)  We will now show that $\Cov{C_i, C_j|A}$ and
$\Var{C_i|A}$ are both $O(\epsilon_n^2) + O(\delta(n, \epsilon_n))$, which,
under these conditions, is polynomial in $1/n$.

We need to modify one more definition from the stochastic block model case: we
re-define $G_n$ as the indicator for the event that $\max_{i \in 1:n}{\|C_i -
\hat{C}_i\| < \epsilon_n}$. (Thus $G_n=1$ with probability $1-\delta(n,
\epsilon_n)$.)  With this in place, we can now proceed much as in Lemma
\ref{lemma:covariance-for-block-models}: by the law of total covariance,
\begin{eqnarray*}
  \lefteqn{\Cov{C_i, C_j|A}} & &\\
  & = & \Expect{\Cov{C_i, C_j|A,G_n}|A} + \Cov{\Expect{C_i|A, G_n},
  \Expect{C_j|A, G_n}|A}.
\end{eqnarray*}

If $G_n=1$, then $C_i = \hat{C}_i + O(\epsilon_n)$ and $C_j = \hat{C}_j +
O(\epsilon_n)$, consequently $\Cov{C_i, C_j|A, G_n=1} = O(\epsilon_n^2)$. If, on
the other hand, $G_n=0$, we do not have such nice control over the covariance of
the true locations, but the fact that they lie in a compact set means that there
is an upper bound, independent of $n$, on the magnitude of their covariance. So
we have shown that
\begin{eqnarray}
  \nonumber \lefteqn{ \Expect{\Cov{C_i, C_j|A,G_n}|A} } & &\\
  \nonumber & = & O(1-\delta(n, \epsilon_n))O(\epsilon_n^2) + O(\delta(n,
  \epsilon_n))O(1)\\
  & = & O(\epsilon_n^2) + O(\delta(n, \epsilon_n)).
  \label{eqn:asta-conditional-covariance}
\end{eqnarray}

Turning to the conditional expectations,
\begin{equation*}
  \Expect{C_i|A, G_n=1} = \hat{C}_i + O(\epsilon_n)
\end{equation*}
and we may define
\begin{equation*}
  \tilde{C}_i \equiv \Expect{C_i|A, G_n=0}
\end{equation*}
which is a function of $A$, and takes values in the convex hull of the compact
set which supports the distribution of $C_i$. Thus
\begin{equation*}
  \Expect{C_i|A, G_n} = G_n\hat{C}_i + G_nO(\epsilon_n) + (1-G_n)\tilde{C}_i.
\end{equation*}
Continuing to imitate the proof of Lemma \ref{lemma:covariance-for-block-models},
\begin{eqnarray*}
  \lefteqn{\Cov{\Expect{C_i|A, G_n}, \Expect{C_j|A, G_n} | A}} & &\\
  & = & \Cov{ G_n\hat{C}_i + G_nO(\epsilon_n) + (1-G_n)\tilde{C}_i, G_n\hat{C}_j +
  G_n O(\epsilon_n) + (1-G_n)\tilde{C}_j|A}\\
  & = & \Var{G_n|A}(\hat{C}_i \hat{C}_j^T + \hat{C}_i O(\epsilon_n) +
  O(\epsilon_n)\hat{C}_j^T + O(\epsilon_n^2))\\
  & & + \Var{1-G_n|A}\tilde{C}_i\tilde{C}_j^T\\
  & &  \Cov{G_n, 1-G_n|A}(\hat{C}_i \tilde{C}_j^T + O(\epsilon_n) \tilde{C}_j^T
  + \tilde{C}_i \hat{C}_j^T + \tilde{C}_i O(\epsilon_n)).
\end{eqnarray*}
By an argument just like the one used in Lemma
\ref{lemma:covariance-for-block-models},  $\Var{G_n|A} = \Var{1-G_n|A} =
O(\delta(n, \epsilon_n))$, and likewise $\Cov{G_n, 1-G_n|A} = O(\delta(n,
\epsilon_n))$. On the other hand, $\hat{C}_i$ and $\tilde{C}_i$ are both $O(1)$.
Thus
\begin{eqnarray}
  \nonumber \lefteqn{\Cov{\Expect{C_i|A, G_n}, \Expect{C_j|A, G_n} | A} } & &\\
  \nonumber & = & O(\delta(n, \epsilon_n)) + O(\epsilon_n \delta(n, \epsilon_n))
  + O(\epsilon_n^2 \delta(n, \epsilon_n))\\
  & = & O(\delta(n, \epsilon_n)) \label{eqn:asta-conditional-expectations}
\end{eqnarray}
since $\epsilon_n \rightarrow 0$.

Combining \eqref{eqn:asta-conditional-covariance} with
\eqref{eqn:asta-conditional-expectations},
\begin{eqnarray}
  \nonumber \lefteqn{ \Cov{C_i, C_j|A} } & &\\
  \nonumber & = & \Expect{\Cov{C_i, C_j|A,G_n}|A} + \Cov{\Expect{C_i|A, G_n},
  \Expect{C_j|A, G_n}|A} \\
  \nonumber & = & O(\epsilon_n^2) + O(\delta(n, \epsilon_n)) + O(\delta(n,
  \epsilon_n))\\
  & = & O(\epsilon_n^2) + O(\delta(n, \epsilon_n)).
  \label{eqn:asta-total-covariance}
\end{eqnarray}

A careful inspection of the preceding steps show that none of them assumed that
$i \neq j$. We may therefore conclude that
\begin{equation*}
  \Var{C_i|A} = \Cov{C_i, C_i|A} = O(\epsilon_n^2) + O(\delta(n, \epsilon_n))
\end{equation*}

Since the bias is $O(\Cov{C_i, C_j|A})+O(\Var{C_i|A})$, the bias is
$O(\epsilon_n^2) + O(\delta(n, \epsilon_n))$. At the corresponding part of the
proof of Theorem \ref{theorem:asymptotic-consistent-for-block-models}, we had a
bias that was $O(\delta(n))$, and an invocation of the GMZZ conditions showed
that this must be exponentially small in $n$. Here, we need to show that
$\epsilon_n^2 \rightarrow 0$ and that $\delta(n, \epsilon_n) \rightarrow 0$ as
well. Invoking the Asta conditions lets us say that
\begin{equation*}
  \delta(n, \epsilon_n) \leq \mathcal{N}(n, \epsilon_n) exp{(-\kappa \epsilon_n
  n^2)}
\end{equation*}
so it's enough to have the right-hand side of this equation approaching zero.
Since the function $\mathcal{N}(n, \epsilon)$ is polynomial in $n$ and
$1/\epsilon$, we can say that
\begin{equation*}
  \log{\delta(n, \epsilon)} = O(\log{n} - \log{\epsilon_n} + \epsilon_n n^2)
\end{equation*}
From this, it's clear that so long as $\epsilon_n n^2 \rightarrow \infty$ at
some polynomial rate, $\delta(n, \epsilon_n)$ will be exponentially small in
some power of $n$, and $\Cov{C_i, C_j|A}$ will be dominated by the
$O(\epsilon_n^2)$ term, which will be polynomial in $n$.

In particular, if $\epsilon_n \propto n^{-r}$, for $0 < r < 2$, then
$\mathcal{N}(n,\epsilon_n)$ is still polynomial in $n$, but $\exp{(-\kappa
\epsilon_n n^2)} = \exp{(-\kappa^{\prime} n^{2-r})}$, so over-all
$\delta(n,\epsilon_n)$ goes to zero exponentially fast in some power of $n$.
Thus we can get $\Cov{C_i, C_j|A} = O(n^{-2r})$ for any $r < 2$.

Having established that both $\Cov{C_i, C_j|A}$ and $\Var{C_i|A}$ are, at most,
$O(n^{-2r})$, reasoning as in the proof of Theorem
\ref{theorem:asymptotic-consistent-for-block-models} shows that the bias, too,
is $O(n^{-2r})$, for some $r < 2$.
\end{proof}

{\em Note:} Attempting to optimize the rate at which $\Cov{C_i, C_j|A}
\rightarrow 0$, by differentiating \eqref{eqn:asta-total-covariance} with
respect to $\epsilon_n$ and setting the derivative to zero, leads to an
un-illuminating transcendental equation, which we omit, because the over-all
convergence rate is still polynomial in $n$.

\subsection*{Acknowledgments}

We thank Andrew C. Thomas, David S. Choi, and Veronica Marotta for many valuable
discussions on these and related ideas over the years. We thank Dena Asta and
Hannah Worrall, for sharing \citet{Asta-thesis} and \citet{Worrall-homophily},
respectively; Chao Gao, Zongming Ma, Anderson Y. Zhang, and Harrison H. Zhou for
sharing code related to \citet{Gao-sbm-minimax_algo}; Oleg Sofrygin for
assistance with simulations using \citet{Sofrygin-simcausal}; and Max Kaplan for
related programming assistance. CRS was supported during this work by grants
from the NSF (DMS1207759 and DMS1418124) and the Institute for New Economic
Thinking (INO1400020) and EM was supported during this work by a grant from
Facebook (Computational Social Science Methodology Research Awards).

\bibliography{control-by-community-discovery}
\bibliographystyle{plainnat}

\end{document}